\documentclass[letterpaper,12pt]{article}
\usepackage{authblk}
\usepackage{multirow, booktabs, graphicx, xcolor,titlesec}
\usepackage[hidelinks]{hyperref}
\usepackage{amsmath, amssymb, amsthm, braket, bbm, physics}

\newcommand{\be}{\begin{equation}}
\newcommand{\ee}{\end{equation}}
\newcommand{\bea}{\begin{eqnarray}}
\newcommand{\eea}{\end{eqnarray}}

\begin{document}

\title{Some aspects of Affleck-Kennedy-Lieb-Tasaki models: tensor network, physical properties, spectral gap, deformation, and quantum computation}
\author[1]{Tzu-Chieh Wei}
\author[2,3]{Robert Raussendorf}
\author[2,3]{Ian Affleck}
\affil[1]{C. N. Yang Institute for Theoretical Physics and
Department of Physics and Astronomy, State University of New York at
Stony Brook, Stony Brook, NY 11794-3840, USA}
\affil[2]{Department of Physics and Astronomy, University of British Columbia, Vancouver, BC V6T 1Z1, Canada}
\affil[3]{Stewart Blusson Quantum Matter Institute, University of British Columbia, Vancouver, BC, Canada}
\date{\today}

\maketitle
\tableofcontents
\newpage
\begin{abstract}
   Affleck, Kennedy, Lieb, and Tasaki constructed a spin-1 model that is isotropic in spins and possesses a provable finite gap above the ground state more than three decades ago.  They also constructed models in two dimensions. Their construction has impacted subsequent research that is still active. In this review article, we review some selected the progresses, such as magnetic ordering of the AKLT models, emerging phases under deforming the AKLT Hamiltonians, symmetry-protected topological order in several AKLT models, their spectral gap, and applications for quantum computation.  
\end{abstract}
\section{Introduction}
The Affleck-Kennedy-Lieb-Tasaki (AKLT) model~\cite{AKLT} gave important confirmation of the Haldane conjecture~\cite{Haldane83,Haldane83b} via an exactly solvable model which can be shown to have an excitation gap and exponentially 
decaying correlation functions. The simplest example is for a spin-1 chain. The Hamiltonian is:
\be H_{\rm AKLT}^{S=1}=\sum_jP^{S=2}(\vec S_j+\vec S_{j+1})=\frac{1}{ 24}\sum_j(\vec S_j+\vec S_{j+1})^2\cdot [(\vec S_j+\vec S_{j+1})^2-2I].
\ee
where $P^{S=2}$ denotes projection onto spin-2, and we shall for convenience denote $P_{j,j+1}^{S=2}\equiv P^{S=2}(\vec S_j+\vec S_{j+1})$. Using, $\vec{S}_j\cdot \vec{S}_j=2$ for spin-1, this can be re-written as
\be H_{\rm AKLT}^{S=1}=\sum_j P_{j,j+1}^{S=2}=\frac{1}{2}\sum_j[\vec S_j\cdot \vec S_{j+1}+(1/3)(\vec S_j\cdot \vec S_{j+1})^2+2/3].
\ee
The ground state must not have a spin-2 state for any pair of neighbouring spins. 
The simplest way of visualizing the ground state $|\Psi_{\rm AKLT}\rangle$ is to decompose the spin-1 into 2 spin-1/2's which are combined into the spin-1 state. A pair of spin-1/2's (or qubits) are then combined into a singlet state on every link as 
sketched below. (We call these valence bonds.) This implies that on two neighboring sites the net spin can only be 0 or 1 and hence such a wave function is annihilated  any  term $P_{j,j+1}^{S=2}$ in the above Hamiltonian: $P_{j,j+1}^{S=2}|\Psi_{\rm AKLT}\rangle=0$. That the ground satisfies the lowest possible energy of each term in the Hamiltonian is called being frustration-free.
For $N$ sites 
with open boundary conditions, the ground state $|\Psi_{\rm AKLT}\rangle$ can be written explicitly as
\begin{eqnarray} 
\label{eq:1dAKLT}|\Psi_{\rm AKLT}\rangle&=&\epsilon^{\alpha_2\alpha_3}\epsilon^{\alpha_4,\alpha_5}...\epsilon^{\alpha_{2n},\alpha_{2n+1}}...\epsilon^{\alpha_{2N-2}\alpha_{2N-1}}\\
&&\quad|\{\alpha_1,\alpha_2\};\{\alpha_3,\alpha_4\};...;\{\alpha_{2n-1}\alpha_{2n}\};...;\{\alpha_{2N-1}\alpha_{2N}\}\rangle,\nonumber
\end{eqnarray}
where $\alpha_i=0,1$ is used to denote the two levels of a spin-1/2 entity,  repeated indices are summed over, and $|\{\alpha_{2k-1},\alpha_{2k}\}\rangle\equiv(|\alpha_{2k-1}\rangle\otimes|\alpha_{2k}\rangle+ |\alpha_{2k}\rangle\otimes |\alpha_{2k-1}\rangle)/\sqrt{2}$ denotes an un-normalized triplet state formed by two virtual qubits on the same physical site $k$. Note that $\alpha_1$ and $\alpha_{2N}$ are uncontracted. This implies effective $S=1/2$ degrees of freedom at the two ends of the chain.  These can be combined 
into a singlet or triplet state which are degenerate for the AKLT model.  It turns out that the basic Heisenberg model also has $S=1/2$ edge states although they have a coupling which drops off exponentially 
with the system size.  If a spin-1 chain contains some random spin=0 defects then pairs of $S=1/2$ states occur on each side of the defect~\cite{Kennedy}. If they are weakly coupled together (but much more strongly coupled 
than between the 2 edges of each chain) then we get random $S=0$ and $S=1$ states at each defect. These have been observed~\cite{edge}. Experimental confirmation of the Haldane gap in spin-1 chains 
was provided by Buyers et al.~\cite{Buyers} and  Renard et al.~\cite{Renard} in two different quasi-one dimensional materials.

\begin{figure}[h]
\centering 
\includegraphics[width=0.7\textwidth]{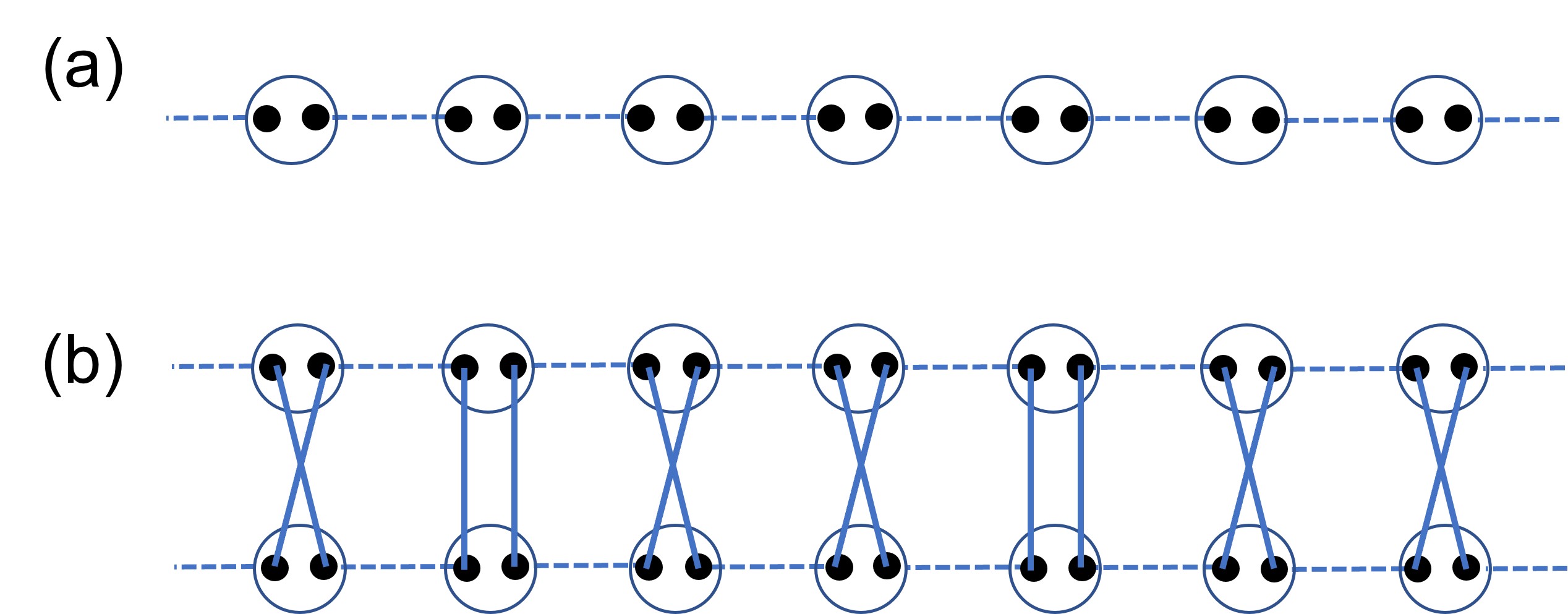}
\caption{(a) Sketch of the $S=1$ AKLT chain. (b) One example term that contributes to the norm square of the AKLT state: $\langle \Psi_{\rm AKLT}|\Psi_{\rm AKLT}\rangle$. The two parallel vertical lines and the two cross lines connecting the upper and lower sites come from, respectively, the first and second term in the expression: $\langle \{\alpha,\beta\}|\{\gamma,\delta\}\rangle=\delta_{\alpha\gamma}\delta_{\beta\delta}+\delta_{\alpha\delta}\delta_{\beta\gamma}$. Each dashed line represents an antisymmetric tensor arising from the singlet shared between neighboring virtual qubits. The overlap is a sum of all possible terms. 
}
\label{AKLT}
\end{figure}
This model has various generalizations. For a spin chain with spin $S$ (an integer) we may form $n$ valence bonds on every link where $n=S$. This is the ground state of the Hamiltonian:
\be H=\sum_j\sum_{S'=1}^S\alpha_{S'}P^{S'}(\vec S_j+\vec S_{j+1}),
\ee
where $\alpha_{S'}>0$. In the rest of the article, we will mostly be concerned with the AKLT model defined in the original paper, i.e., the magnitude  of the spin $S$ at a site is determined by the number $z$ of its neighbors: $S=z/2$.

The exact ground state correlation function was calculated for the $S=1$ case in \cite{AKLT} and is
\be \langle 0|S^\alpha_jS^\beta_k|0\rangle=\frac{4}{ 3}(-1)^{k-j}3^{-|k-j|}\delta_{\alpha\beta}.\ee
We shall see below the alternative approach using matrix product states (MPS)~\cite{Fannes1992,MPS} other than that in the original work. 
Such an exponential decay in the correlation function suggests the existence of a gap. A
rigorous proof of the gap between the ground state and first excited state for  periodic boundary conditions was also given in~\cite{AKLT}. (We shall also see other approaches for one- and two-dimensions below.) Therefore, this spin-1 model has $SO(3)$ rotational symmetry and possesses a unique magnetically disordered  ground state (in the thermodynamic limit) and a nonzero energy gap.  This model is recognized as an example of symmetry-protected topological order (SPTO)~\cite{Gu,Pollmann,Chen}. It is manifested in the fractionalization (from spin-1 to spin-1/2)  of gapless excitations at the boundary. Their response to the symmetry action is $SU(2)$, a projective representation of the original $SO(3)$.
We refer the readers to the review article~\cite{AffleckReview} on the relation of the  Haldane gap to the vanishing of a topological theta term with $\theta=2\pi s$, i.e. equivalent to zero for integer $s$ spins, and the Lieb-Schultz-Mattis theorem for half-odd integer $s$ to the existence of the $\theta$ term in the nonlinear sigma model, whose Lagrangian is 
\be
\label{eq:NLsigma}
{\cal L}=\frac{1}{2g}\partial_\mu \vec{n}\cdot \partial^\mu\vec{n} + \frac{\theta}{8\pi}\epsilon^{\mu\nu}\vec{n}\cdot \partial_\mu \vec{n}\times \partial_\nu \vec{n},
\ee
where $\vec{n}(x,t)$ is the vector order parameter with
unit length.

\begin{figure}
\centering 
\includegraphics[width=0.75\textwidth]{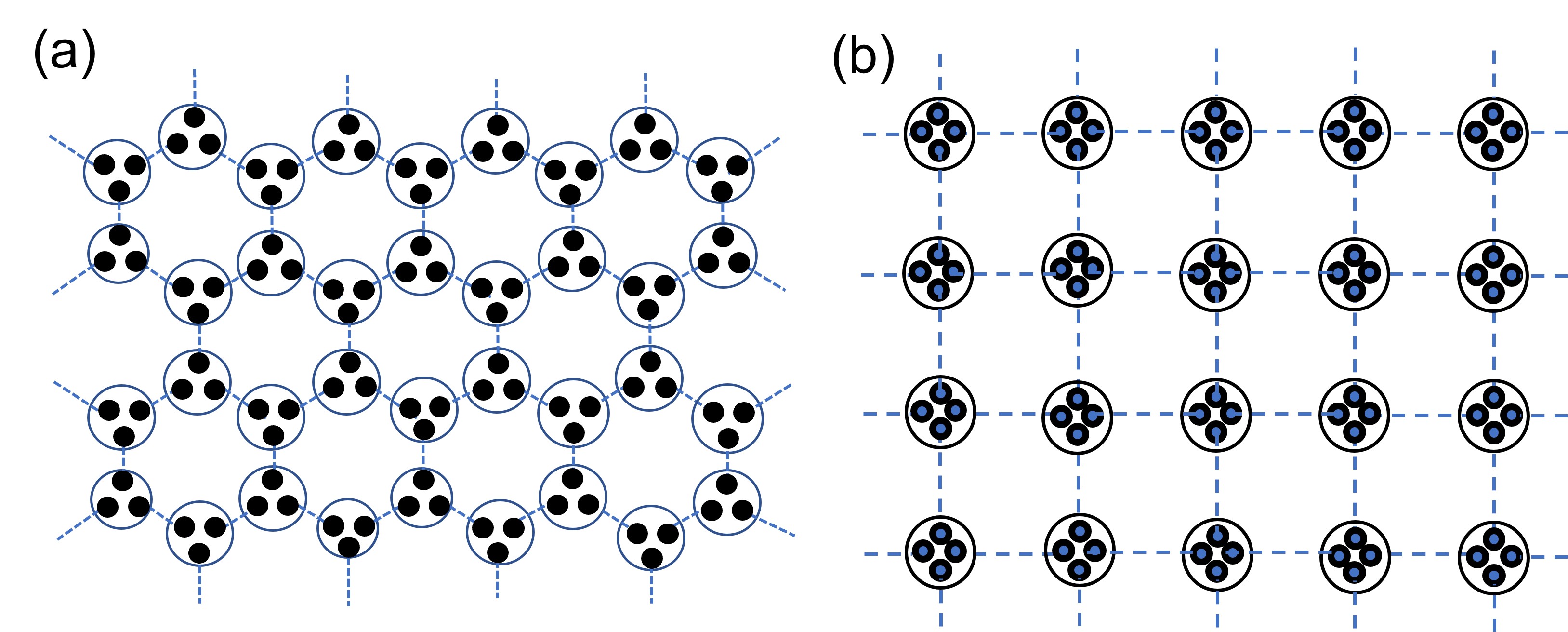}
\caption{\label{fig:HoneycombSquare}Sketch of valence bond construction for (a) the $S=3/2$ AKLT state on the honeycomb lattice and (b) the $S=2$ AKLT state on the square lattice. }

\end{figure}

The AKLT states can be extended to two and three dimensions. The simplest extension, relevant for the quantum computing applications discussed below, is the honeycomb lattice with spin-3/2. Each 
site has 3 nearest neighbours.  We decompose the spin-3/2 into 3 spin-1/2's and form a valence bond on every link, as illustrated in Fig.~\ref{fig:HoneycombSquare}. The above 1D spin-1 wave function provides one of the earliest examples of matrix-product states and the 2D spin-3/2 AKLT wave function is an example of  projected  entangled pair states (PEPS)~\cite{PEPS}.  These tensor-network representations turn out to give a useful tool. For example, the wave function~(\ref{eq:1dAKLT}) can normalized is a diagrammatic way, using the property of the triplet $\langle \{\alpha,\beta\}|\{\gamma,\delta\}\rangle=\delta_{\alpha\gamma}\delta_{\beta\delta}+\delta_{\alpha\delta}\delta_{\beta\gamma}$, and observables and correlation functions can also be evaluated by summing various diagrams; see e.g. Fig.~\ref{AKLT}.  By employing MPS, these can be evaluated using elementary transverse matrix calculation.

An alternative way of studying the AKLT states was developed in~\cite{AAH}.  We may introduce 2 bosons on every site, $a$ and $b$ with:
\bea S^z=(1/2)(a^\dagger a-b^\dagger b), \
S^+=a^\dagger b, \
S^-=b^\dagger a.
\eea
The AKLT state, for a general lattice, then becomes:
\be \label{eqn:boson}|\psi \rangle=\prod_{ij}
(a^\dagger_ib^\dagger_j-b^\dagger_ia^\dagger_j)^M|0\rangle,
\ee
where $M$ is the number of valence bonds on each link. The wave function by using the coherent state representation gives rise to a classical partition function of antiferromagnets. The magnetic ordering of the corresponding AKLT state can be studied via the classical partition function.

The remaining structure of this book chapter is as follows. In Sec.~\ref{sec:TN}, we describe AKLT states using tensor-network representations, including matrix-product states MPS and PEPS. In Sec.~\ref{sec:magnetic}, we the magnetic ordering of AKLT models. In Sec.~\ref{sec:SPT}, we describe some understanding of symmetry-protected topological order in several AKLT states. In Sec.~\ref{sec:hidden}, we describe certain hidden orders in AKLT states; one of this is related to symmetry-protected topological order and the other is related to cluster states and is useful for quantum computation. In Sec.~\ref{sec:MBQC}, we give an explanation how AKLT states can be used for quantum computation, in particular, via the scheme of measurement-based quantum computation. AKLT states are among a few spin systems being explored for such a  measured-based approach~\cite{Wei2018}. In Sec.~\ref{sec:gap}, we explain techniques that lead to rigorous establishment of some two-dimensional AKLT models, such as the one on the  hexagonal lattice. In Sec.~\ref{sec:deformed}, we discuss the scenario beyond the AKLT models by deforming them locally. We conclude this chapter in Sec.~\ref{sec:conclusion}.
\section{Tensor-network picture: MPS and PEPS}
\label{sec:TN}
A modern perspective of AKLT states is that they can be represented by tensor-network states, such as the matrix-product states (MPS) in one dimension and the projected entangled pair states (PEPS) in one and higher dimensions. To describe these states one places certain number of virtual qudits on each lattice sites according to the lattice coordination number, and the two qudits  associated with an edge form a maximally entangled states. Then one maps the Hilbert space of the qudits on a site to that of a physical spin. 
\subsection{
1D AKLT chain}

 Each virtual qubit is entangled with a virtual qubit on its neighboring site in the form of a spin-singlet (un-normalized and conveniently expressed in a product of a row vector of kets with a column vector of kets): 
\begin{equation}
|01\rangle-|10\rangle = \left(\begin{array}{cc}
|0\rangle & |1\rangle \end{array} \right) \left(\begin{array}{c}
|1\rangle \\
-|0\rangle
\end{array}\right),
\end{equation}
where the virtual qubit on the right side of a site is represented by the row vector and the one on the left side of the next site is represented by a column vector.
Combining the two virtual qubits on each site, we have
\begin{equation}
\left(\begin{array}{c}
|1\rangle \\
-|0\rangle
\end{array}\right)
\left(\begin{array}{cc}
|0\rangle & |1\rangle \end{array} \right) =
\left(\begin{array}{cc}
|10\rangle & |11\rangle \\
-|00\rangle & -|01\rangle\end{array} \right),
\end{equation}
which is the local matrix whose product represents pairs of singlets, with the boundary condition unspecified. 

\begin{figure}
\centering 
\includegraphics[width=0.7\textwidth]{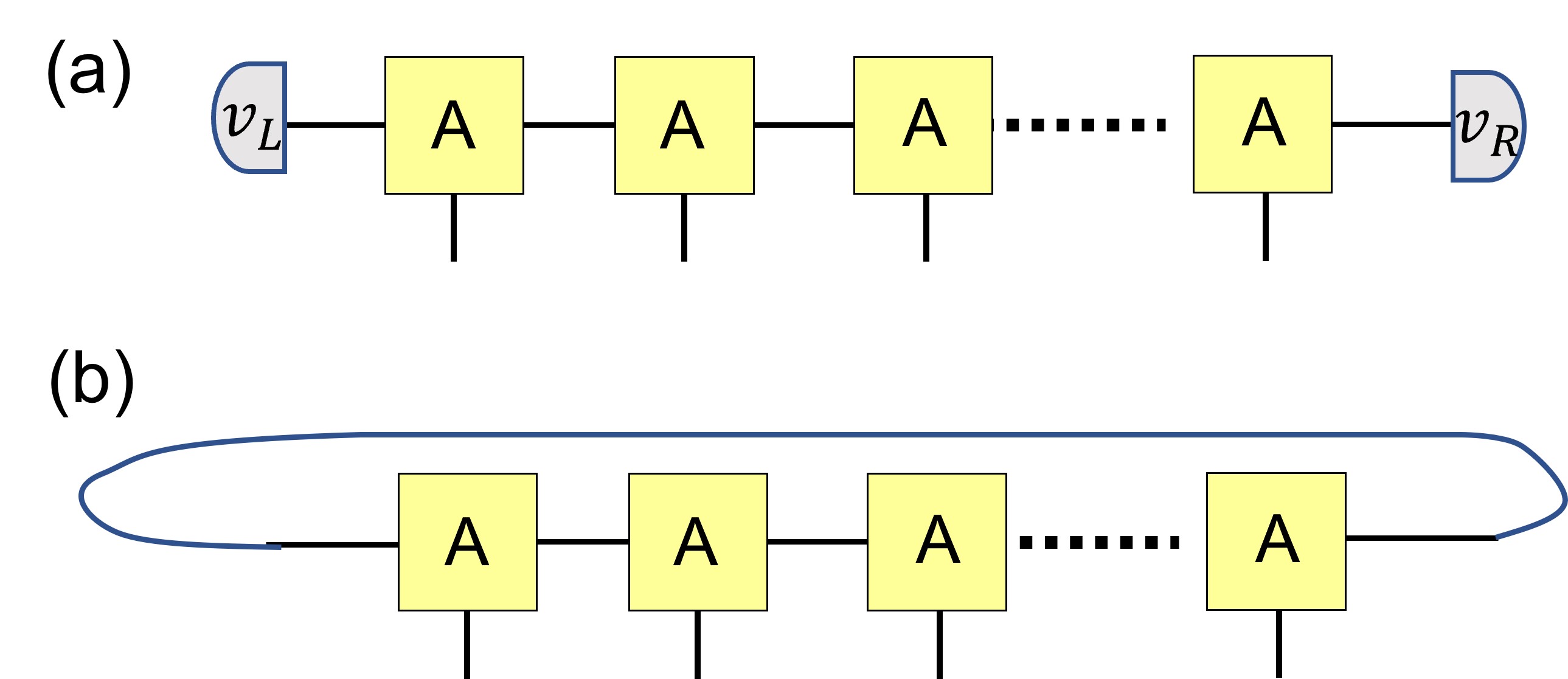}
\caption{Schematic of matrix-product states: (a) open boundary condition and (b) periodic boundary condition. }
\label{fig:MPS}
\end{figure}
The local mapping from two virtual qubits to a single spin 1 (with basis states $|S=1,S_z=+1\rangle, |S=1,S_z=0\rangle, |S=1,S_z=-1\rangle$) is given by (omitting the $S$ and $S_z$ labels) \begin{equation} P_v=|+1\rangle \langle 00| +|0\rangle (\langle 01|+\langle10|)/\sqrt{2} +|-1\rangle \langle 11|,
\end{equation}
where $|0\rangle$ on the second term is $|S=1,S_z=0\rangle$.
The action of $P_v$ on the two virtual qubits yields
\begin{equation}
\label{eqn:MPSa}
P_v
\left(\begin{array}{cc}
|10\rangle & |11\rangle \\
-|00\rangle & -|01\rangle\end{array} \right)=\left(\begin{array}{cc}
|0\rangle/\sqrt{2} & |-1\rangle \\
-|+1\rangle & -|0\rangle\sqrt{2}\end{array} \right)=|0\rangle \frac{1}{\sqrt{2}}\sigma_z + |+1\rangle (-\sigma^-) + |-1\rangle \sigma^+.\end{equation}
Thus, we derive the three matrices corresponding to the three physical degrees $|0\rangle$, $|+1\rangle$, and $|-1\rangle$, i.e.
\begin{equation}
\label{eqn:1DMPS}
 A_{0}=\sigma_z/\sqrt{2}, \ A_{+1}=-\sigma^-, \ A_{-1}=\sigma^+.   
\end{equation} 
These matrices describe the system in the bulk and one can specify the boundary condition. For example, for the open boundary condition, we can specify a left and right vectors $\vec{v}_{L/R}$ applied to the product of matrices:
\begin{equation}
\label{eqn:1Dopen}
    |\psi_{\rm open}\rangle= \sum_{s=0,\pm 1} v_L^{T} A_{s_1}A_{s_2}\cdots A_{s_N} v_R |s_1,s_2,\dots,s_N\rangle,
\end{equation}
which represents the ground state of the spin-1 AKLT chain with open boundary (i.e. the first spin is not coupled to the last spin $H=\sum_{j=1}^{N-1}P_{j,j+1}^{S=2}$); see Fig.~\ref{fig:MPS}a. The boundary spin-1/2 degrees of freedom are seen from the two-component vectors $v_L$ and $v_R$ being arbitrary.  

To describe the ground state of the periodic AKLT chain, we simply take the trace of the matrix product,
\begin{equation}
\label{eqn:1Dperiodic}
    |\psi_{\rm periodic}\rangle= \sum_{s=0,\pm 1} {\rm Tr}( A_{s_1}A_{s_2}\cdots A_{s_N} ) |s_1,s_2,\dots,s_N\rangle.
\end{equation}
This is illustrated in Fig.~\ref{fig:MPS}b. The Hamiltonian such that $|\psi_{\rm periodic}\rangle$ is the ground state is  $H=\sum_{i=1}^{N} P^{[S=2]}_{i,i+1}$, with site $N+1$ identified with site 1.
We will discuss the degeneracy of the ground states, calculations of observables and correlations, and the proof of gap below.

\subsection{Two dimensions}
Here, we review the AKLT states on the hexagonal and square lattices and their Hamiltonians. 
\subsubsection{Honeycomb/hexagonal lattice} 
\begin{figure}[h]
\centering 
\includegraphics[width=0.7\textwidth]{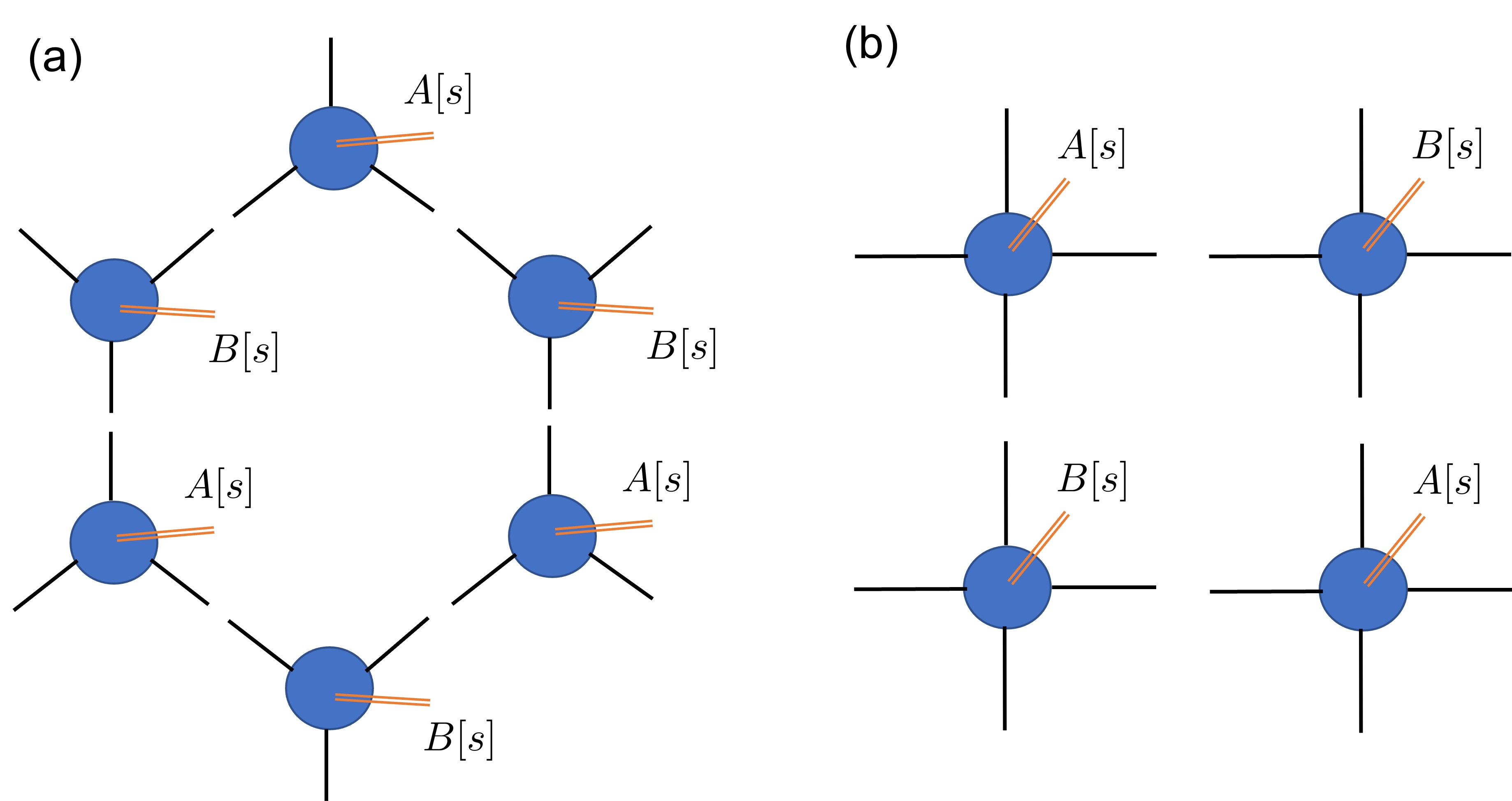}
\caption{Schematic of tensors for AKLT ground states on (a) the hexagonal lattice and (b) the square lattice. The double lines represent physical degrees of freedom, whereas the thin lines represent the virtual indices. For a fixed set of physical indices, the amplitude for the component is proportional to the value given by the tensor contraction of all  virtual indices. }
\label{fig:2DTensors}
\end{figure}

Each site contains three virtual qubits, each forming a singlet with its neighboring virtual qubit; see Fig.~\ref{fig:HoneycombSquare}a. The local projection is from that of three virtual qubits to their symmetric subspace, which is identified as the Hilbert space of a physical spin-3/2 site. The projection is given as
\begin{equation}\label{Proj}
    P_v=|S_z=+\!3/2\rangle \langle 000| + |S_z=-\!3/2\rangle \langle 111| + |S_z=+\!1/2\rangle \langle W| + |S_z=-\!1/2\rangle \langle \bar{W}|,
\end{equation}
where we have defined for convenience
\begin{eqnarray}
 |W\rangle&\equiv& \frac{1}{\sqrt{3}}(|001\rangle + |010\rangle + |100\rangle),\\
 |\bar{W}\rangle&\equiv& \frac{1}{\sqrt{3}}(|110\rangle + |101\rangle + |011\rangle).
\end{eqnarray}
One can generalize the representation of the matrix product states to 2D and in this case is the tensor product. Here we can choose two different types of sites, labelled by A and B, respectively, to write the nonzero components of a tensor corresponding to a physical index $s$ for tensor $A_s$ or $s'$ for tensor $B_{s'
}$. There are three virtual indices for each $A_s$ and $B_{s'}$, whose structure is illustrated in Fig.~\ref{fig:2DTensors}. The nonzero elements in tensor $A$ can be read off from $P_v$ in Eq.~(\ref{Proj}). For example, $A[3/2]_{000}=1$, $A[-3/2]_{111}=1$, etc. Those in tensor $B$ can be obtained from
\begin{eqnarray}
  \label{ProjB} && P_v\,(i\sigma_y)\otimes(i\sigma_y)\otimes(i\sigma_y)\\
    &&=|S_z=+\!3/2\rangle \langle 111| - |S_z=-\!3/2\rangle \langle 000| - |S_z=+\!1/2\rangle \langle {W}| + |S_z=-\!1/2\rangle \langle \bar{W}|. \nonumber
\end{eqnarray}

The parent Hamiltonian can also be straightforwardly obtained from the projector onto the join spin-3 subspace of two neighboring sites $i$ and $j$: $H=\sum_{\langle i, j\rangle} P_{i,j}^{(S=3)}$ as there are 6 virtual qubits with two forming a singlet, indicating that the total spin magnitude cannot exceed $S=2$. Thus, the constructed AKLT state is the ground state of this Hamiltonian composed of a sum of projectors. Translating it to the spin-3/2 operators, we have
\begin{equation}
    H_{\rm AKLT}^{S=3/2}=\sum_{{\rm edge}\,\langle i,j\rangle}\hat{P}_{i,j}^{(S=3)}=\frac{27}{160}\sum_{{\rm edge}\,\langle i,j\rangle}\Big[ \vec{S}_i\cdot \vec{S}_{j}+\frac{116}{243}(\vec{S}_i\cdot \vec{S}_{j})^2+\frac{16}{243}(\vec{S}_i\cdot \vec{S}_{j})^3+\frac{55}{108} \Big].
\end{equation}

The original paper~\cite{AKLT} uses a different representation (i.e. a polymer representation) via links and shows that the correlation function $C(r)$ is bounded above by an exponential decaying function.
\subsubsection{Square lattice}
We refer to see Fig.~\ref{fig:HoneycombSquare}b for the schematic of the construction. For $S=2$ case, the local mapping from 4 virtual qubits to $S=2$ Hilbert space is as follows,
\begin{equation}
    P_v^{[S=2]}=|+\!2\rangle \langle 0000| + |-\!2\rangle \langle 1111| + |+\!1\rangle \langle S(4,1)| + |-\!1\rangle \langle S(4,3)|+|0\rangle\langle S(4,2)|,
\end{equation}
where $|S(n,k)\rangle$ is the Dicke state with superposition of $k$ 1's and $(n-k)$ 0's. Due the singlets along edges, we can choose to have two types of tensors on A and B sublattices, with tensor $A$ being readily read off from $P_v^{[S=2]}$. The tensor on the other sublattice (B) is related to that of $A$ via $P_v^{[S=2]}\sigma_y\otimes \sigma_y\otimes\sigma_y\otimes\sigma_y$.
We note that it is also possible to choose the tensors uniformly for each site, e.g., $P_v^{[S=2]}\sigma_y\otimes \sigma_y\otimes I\otimes I$.

The parent Hamiltonian for $S=2$ AKLT model is obtained from the two-site projector onto the joint $S=4$ subspace,
\begin{eqnarray} \label{eqn:spin2AKLT}
 H^{S=2}_{\rm AKLT}
=\sum_{{\rm edge}\,\langle i,j\rangle}\hat{P}_{i,j}^{(S=4)}=\frac{1}{28}\sum_{\langle i,j\rangle}\Big[ \vec{S}_i\cdot
\vec{S}_{j}+\frac{7}{10}(\vec{S}_i\cdot \vec{S}_{j})^2
 +\frac{7}{45}(\vec{S}_i\cdot \vec{S}_{j})^3 +
\frac{1}{90}(\vec{S}_i\cdot \vec{S}_{j})^4\Big].
\end{eqnarray}
The tensors in the PEPS representation can be read off from $P_v^{[S=2]}$, and the schematic picture is given in Fig.~\ref{fig:2DTensors}b. Similar to previous arguments, the AKLT state above is a ground state of the Hamiltonian~(\ref{eqn:spin2AKLT}). The correlation function in its ground state was shown in Ref.~\cite{KLT} to be bounded by an exponentially decay function.

\subsection{Boundary conditions and degeneracy of AKLT models}
Kennedy, Lieb and Tasaki used the polynomial representation (in terms of `spinors' $u_j$ and $v_j$ see Sec.~\ref{sec:magnetic}) of Arova, Auerbach and Haldane and showed that the AKLT model on any lattice has its ground state wave function written as~\cite{KLT}
\be \Psi=
\Phi\prod_{i,j|\langle i,j\rangle{\rm edge} } (u_i v_j-u_jv_i),
\ee 
where $\Phi$ is a unique polynomial of those $u$'s and $v$'s on the boundary. This means that if there is degeneracy, it can only come from the boundary via  $\Phi$.  In particular, in the periodic boundary condition, $\Phi=1$, and hence, the finite-volume ground state is unique. One may naively think that the infinite-volume limit,  AKLT models have a unique ground state. This would be correct if one can show that there is no N\'eel order or alternatively that the correlation functions are exponentially decaying, as done by KLT~\cite{KLT}. 

Pomata and Wei showed the degeneracy of open boundary condition is related to the number of open legs at the boundary; see Supplemental Materials of Ref.~\cite{PomataWei2020}. In particular, any boundary site that has $k$ dangling virtual qubits (not forming singlets with other sites) contribute to a degeneracy of $k+1$. In terms of the tensor-network description, by symmetrizing these dangling tensors, the resultant tensor that maps from these uncontracted bonds to the degenerate ground state is a bijective tensor. Their proof uses induction by beginning with a disjointed  subgragh  (which is bijective) and then showing that bijectivity is preserved when edges are added.

\section{Magnetic ordering}

\begin{figure}
\centering 
\includegraphics[width=0.75\textwidth]{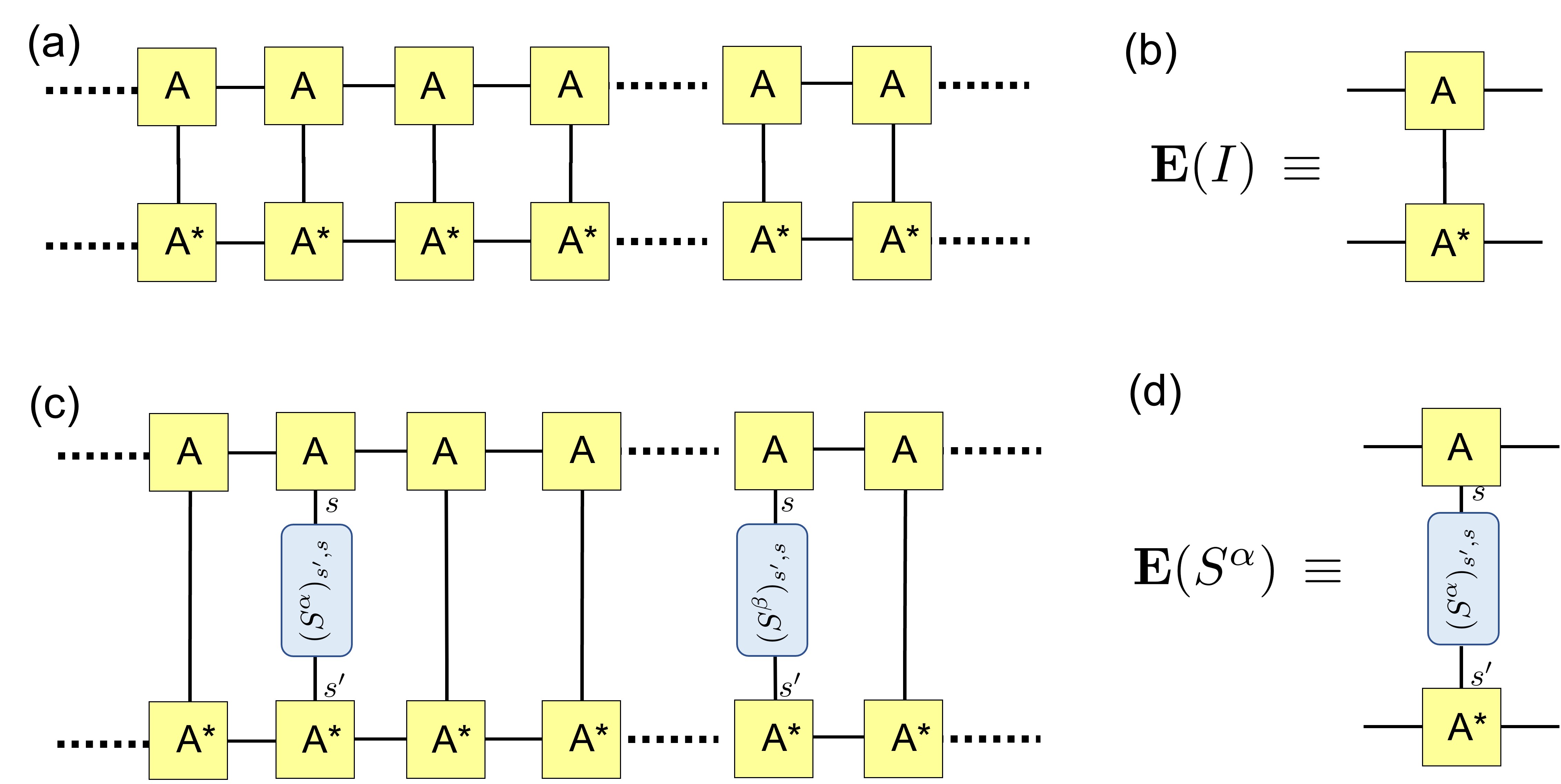}
\caption{Illustration of how to compute expectation values (such as the correlation functions) using the MPS formalism. (a) The diagram represents the normal square of the wavefunction $\langle\psi|\psi\rangle$. (b) The local transfer matrix $\mathbf{E}(I)\equiv \sum_s A[s] \otimes A^*[s]$. (c) The correlation function $\langle\psi|S^\alpha_i S^\beta_{i+r}|\psi\rangle$, which should be normalized by the expression in (a). (d) The local transfer matrix associated with a spin operator $S^\alpha$: $\mathbf{E}(S^\alpha)\equiv  \sum_{s,s'} (S^\alpha)_{s',s} A[s] \otimes A^*[s]$.}
\label{fig:MPSTransferMatrix}
\end{figure}
The valence-bond construction of AKLT states seems to imply that there is no magnetic ordering. It turns out that this issue is slightly complicated as one needs to consider the thermodynamic limit. For the one dimensional AKLT state, it was shown in Ref.~\cite{AKLT} that there is no magnetic ordering, i.e. $\sum_i (-1)^i \langle S_i^\alpha\rangle/N =0$ or $\langle S_i^\alpha S_{i+r}^\alpha\rangle =(-1)^r4/3^{r+1}\rightarrow 0$ as $r\rightarrow \infty$. This can be  calculated using the MPS formalism, illustrated in Fig.~\ref{fig:MPSTransferMatrix}.  We note that in the infinite system limit, one only needs to using the eigenvector corresponding the the largest eigenvalue in magnitude in evaluating the expectation from the left and right boundaries. We leave the details for readers to work out on their own. Although there is no N\'eel order, we do see the weak antiferromangetic correlation from the factor $(-1)^r$.

\begin{figure}
\centering 
\includegraphics[width=0.9\textwidth]{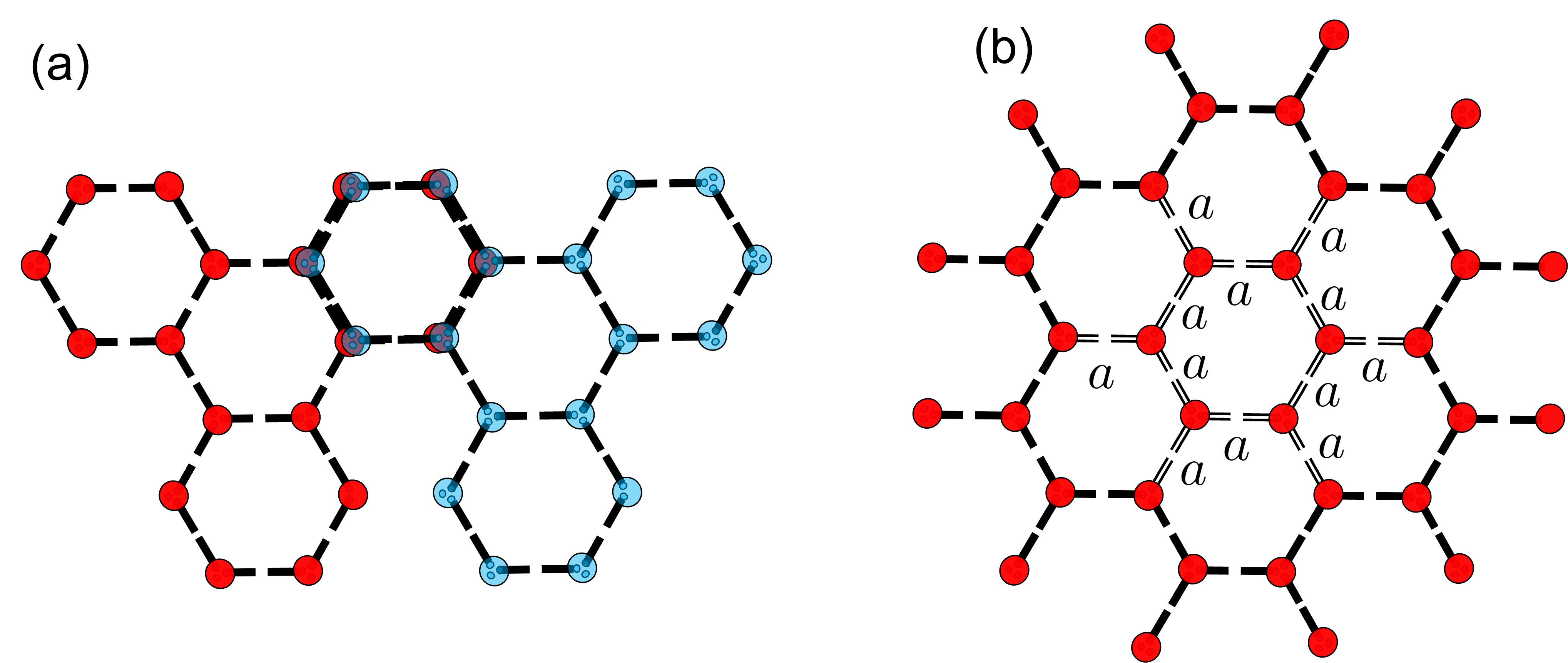}
\caption{Examples of $n=1$ decorated lattices on the original (a) Bethe lattice with $z=3$, (b) honeycomb lattice, and (c) square lattice. Smaller dots represented the inserted $S=1$ sites on every edge of the original lattice.  }
\label{fig:Decorated}
\end{figure}

However, antiferromagnetic ordering does occur on the Bethe lattice (or the Cayley tree) with coordination number $z=5$ or larger as shown in the original work of AKLT~\cite{AKLT}. It seems that $z_c=4$ is the critical coordination number. We note that recently Pomata  considered decorating each edge in the Bethe lattice by adding $n$ spin-1 sites (i.e. a spin-1 chain with $n$ sites; see e.g. Fig.~\ref{fig:Decorated}a) and showed that the critical coordination number $z_c(n)=3^{n+1}+1$~\cite{Pomata21}. This is consistent with the picture that smaller spin-$S$ has larger quantum fluctuations than large spin-$S$;  for the coordination number $z$, the spin magnitude is $S=z/2$ and decoration of $n$ $S=1$ sites on each edge pushes the ordering to occur at a larger coordination $z_c(n)$. 

One useful approach to tackle the issue of ordering is to use the Schwinger-boson representation by Arovas, Auerbach and Haldane and consider the wavefunction in the coherent-state basis $
|\hat{n}\rangle=\frac{1}{\sqrt{(2S)!}} ( u \hat{a}^\dagger + v \hat{b}^\dagger)^{2S}|{\rm vacuum}\rangle.$ The AKLT wavefunction becomes $\Psi(\{u,v\})=\langle \{\hat{n}\}|\psi_{\rm AKLT}\rangle= \prod_{\langle i,j\rangle} (u_iv_j-u_jv_i)^M$, where $M$ is the number of singlets on an edge (which is 1 for the original AKLT states). One maps the norm square of the wavefunction  to a classical O(3) antiferromagnetic model, i.e.,
\begin{equation}
    \Psi^*\Psi= e^{-H_{\rm cl}(\{\hat{n}\})/T},
\end{equation}
where $T=1/M$ and \begin{equation}
    H_{\rm cl}= -\sum_{\langle i,j\rangle}\ln \frac{1-\hat{n}_i\cdot \hat{n}_j}{2}.
\end{equation}
This is essentially an antiferromagnetic interaction, as one can see from expanding the logarithm: $H_{\rm cl}\sim \sum_{\langle i,j\rangle} \hat{n}_\cdot \hat{n}_j -(\hat{n}_\cdot \hat{n}_j)^3/3+\dots$. On a bipartite lattice, this is equivalent to a ferromagnetic model by setting $\hat{n}'_j=(-1)^j \hat{n}_j$. 

Using the Mermin-Wagner theorem, one readily sees that there is no magnetic ordering for AKLT models on 1D and 2D regular lattices~\cite{Param}. However, it was shown by Monte Carlo simulations that there is an antiferromagnetic ordering for the AKLT model on the 3D cubic lattice, but not on the 3D diamond lattice.  The ordering implies spontaneous symmetry breaking and shows that the AKLT model has ground-state degeneracy greater than one on the cubic lattice in the thermodynamic limit~\cite{Param}. 
\label{sec:magnetic}
\section{Symmetry-Protected Topological Order}
\label{sec:SPT}
In this section, we examine several AKLT models from the perspective of symmetry-protected topological order.
\subsection{SPT order of 1D AKLT state} 

AKLT is a symmetry-protected  topological (SPT) state, e.g. by $Z_2\times Z_2$ symmetry (generated by rotation around x or z by 180 degrees), a discrete subgroup of $SO(3)$.
We can examine the action of these group elements on the local matrices of MPS. The symmetry group is generated by the two rotations on the physical spin basis ($|x\rangle$, $|y\rangle$, and $|z\rangle$),
\begin{equation} U_x(\pi)=\left( \begin{array}{ccc}
0& 0 & -1\\
0 & -1 & 0\\
-1& 0 & 0\end{array}\right), \  U_y(\pi)=\left( \begin{array}{ccc}
0& 0 & 1\\
0 & -1 & 0\\
1& 0 & 0\end{array}\right), \
U_z(\pi)=\left( \begin{array}{ccc}
-1 & 0 & 0 \\
0 & 1 & 0\\
0 & 0 & -1 \end{array}\right).\end{equation} 
By directly applying their actions on the local matrices in Eq.~(\ref{eqn:1DMPS}), we have, for examples,
\begin{equation} \sum_{\beta}[U_x(\pi)]_{\alpha,\beta}A_\beta = \sigma_x \cdot A_\beta \cdot \sigma_x, \ \ \sum_{\beta}[U_z(\pi)]_{\alpha,\beta}A_\beta = \sigma_z \cdot A_\beta \cdot \sigma_z.\end{equation}
We see that $\sigma_x$, $\sigma_y$, and $\sigma_z$ form  a projective representation of $U_x$, $U_y$, and $U_z$, respectively.  They generate the  single-qubit Pauli group and represent the symmetry action on the boundary degree of a semi-infinite chain, which exhibits fractionalization. Thus, the 1D AKLT state exhibits a nontrivial SPT order~\cite{Gu,Pollmann,Chen}. Such an SPT order can also be detected by the string order parameter~\cite{denNijs,KennedyTasaki}; see Eq.~(\ref{eq:stringorder}) below.

Another approach to understand the SPTO in this 1D AKLT chain is the topological theta term in Eq.~(\ref{eq:NLsigma}) with $\theta=2\pi$, and the ground state wave function of the SPT phase described the nonlinear sigma model can be expressed as superposition of these spin vectors decorated by a local phase given by a Wess-Zumino-Witten   term~\cite{XuSenthil},
\be
|\Psi\rangle\sim \int_{S^d} d^dx\, e^{-\frac{1}{g} (\nabla \vec{n})^2-{\rm WZW}_d[\vec{n}]}|\vec{n}(x)\rangle,
\ee
where the 1D WZW term is related to the theta term in~\ref{eq:NLsigma},
\be
 {\rm WZW}_1[\vec{n}]=\int_{0}^1 du\,\frac{i2\pi}{8\pi}\epsilon^{\mu\nu}\vec{n}\cdot \partial_\mu \vec{n}\times \partial_\nu \vec{n}, \ \ \mbox{with $\mu,\nu=x,u$},
\ee
where $u$ extends the space to an additional dimension. Based on this picture, You et al. propose a strange correlator~\cite{YouBiRasmussenSlagleXu2014},
\be
C(\vec{r},\vec{r}')=\frac{\langle \Omega |\phi(\vec{r})\phi(\vec{r}') |\Psi\rangle}{\langle \Omega |\Psi\rangle}\ee
for some local operator $\phi(\vec{r})$, to detect the presence of SPTO in the state of concern $|\Psi\rangle$, where $|\Omega\rangle$ is an product trivial state. According to You et al. this strange correlator for SPT states is either constant or polynomially decaying. Using the MPS formalism illustrated in Fig.~\ref{fig:MPSTransferMatrix}, one can easily calculate this and obtain that $C(r,r')=2$
for $|\Psi\rangle$ being the 1D spin-1 AKLT wave function and $|\Omega\rangle=|00...0\rangle$.

\subsection{Two dimensions: honeycomb and square lattices}

Here, we will also examine the symmetry action in terms of virtual degrees of freedom using the PEPS formalism,  we will find that it forms only a projective representation, and this seems to imply weak SPT order for both the 2D AKLT models on both the honeycomb lattice and square lattices. However, the study of the strange correlator shows the difference between the two models, indicating the AKLT state on the square lattice exhibits nontrivial SPTO.

First, the $U_x$, $U_y$ and $U_z$ of rotations around the three respective axes by an angle of $\pi$ in the spin-3/2 representation are
\[ U_x(\pi)=\left( \begin{array}{cccc}
0& 0 & 0 & i\\
0 & 0 & i & 0\\
0 & i & 0 & 0 \\
i & 0 & 0 & 0\end{array}\right), \
U_y(\pi)=\left( \begin{array}{cccc}
0& 0 & 0 & -1\\
0 & 0 & 1 & 0\\
0 & -1 & 0 & 0 \\
1 & 0 & 0 & 0\end{array}\right), \
U_z(\pi)=\left( \begin{array}{cccc}
i & 0 & 0 & 0 \\
0 & -i & 0 & 0\\
0 & 0 & i & 0\\
0 & 0 & 0 & -i\end{array}\right),\] 
which does not give a faithful representation of $Z_2\times Z_2$, as e.g. $U_x(\pi)U_z(\pi)=-U_z(\pi)U_x(\pi)$ and $U_x(\pi)^2=-I$. It is actually a representation of the quaternion group. One can check that, similar to the 1D case, the symmetry action on the physical index can be replaced by action on the virtual indices with Pauli matrices, up to a global phase. This shows that the action on the boundary is at best a projective representation of $Z_2\otimes Z_2$, but for the strong SPTO, the symmetry actions on the boundary need to be a manifestation of third group cohomology group~\cite{Chen2013,ElseNayak}. Hence we conclude that the 2D AKLT state on the honeycomb lattice is only a weak SPT order.  This is confirmed by the strange-correlator calculations by Wierschem and Beach~\cite{WierschemBeach2016}, which display exponential decay.

One can perform a similar analysis for the square-lattice case and find that the symmetry action of $U_\alpha(\pi)$ on the physical index is equivalent to applying Pauli $\sigma_\alpha$ on all four virtual indices, which is a projective representation of $Z_2\otimes Z_2$. This suggests that the AKLT state on the square lattice is also weak SPT ordered in terms of cohomology.  However, the issue of the SPT order for the square-lattice AKLT model is tricky. You et al. calculated the strange correlator for the AKLT state on the square-lattice AKLT state, and found that it  is power-law decaying, as opposed to the exponential decay in the former. This shows that there is  strong SPT order in the 2D AKLT state on the square lattice~\cite{YouBiRasmussenSlagleXu2014}.  

We also mention that a work by Haldane in 1988~\cite{Haldane1988} on an  O(3)  non-linear sigma model study for
 2D quantum Heisenberg antiferromagnets shows that
certain
tunneling processes between states of different topology
 have amplitudes  sensitive to whether the microscopic spin is a half integer, odd integer, or
even integer. The AKLT model on the square lattice is such an example of even-integer $S$, which has a unique disordered ground state, likely with a gap. 

\section{Hidden  order in AKLT states}
\label{sec:hidden}
\subsection{String order parameter}
The 1D AKLT state such as in Eq.~\eqref{eqn:1Dopen} written in the MPS form allows us to see the hidden antiferromagnetic ordering in the state. The components in the wavefunction cannot have two equal $S_z=1$ (or $S_z=-1$) spaced by any number of $S_z=0$. The allowed ones are $...(+1)0...0(-1)...$ or $...(-1)0...0(+1)...$. The spin configuration after stripping off the $0$'s should be antiferromagnetic. This result can be understood by the MPS picture, as $\sigma^\pm (\sigma_z)^n \sigma^\mp =0$. 

This hidden  ordering can also be probed by a string order parameter~\cite{denNijs,KennedyTasaki}
\begin{equation}
\label{eq:stringorder}
    \Pi_{i,i+r}^{\alpha}=S^\alpha_i e^{i\pi\sum_{j=i+1}^{i+r-1}S_j^\alpha}S^\alpha_{i+r}, \ \alpha=x,y,z.
\end{equation}
Den Nijs and
Rommelse   argued
that this order parameter  is  nonzero in the Haldane phase and can be used to distinguish from other gapped phases~\cite{denNijs}.
Kennedy and Tasaki found a nonlocal unitary transformation $U$ that takes $-\Pi^{\alpha}$ to $S^\alpha_i S^\alpha_{i+r}$ and thus the latter detects ferromagnetic ordering in the transformed Hamiltonian $\tilde{H}=UHU^{-1}$. The essential symmetry of concern is the $Z_2\times Z_2$ of $\tilde{H}$ and the Haldane phase corresponds to complete breaking of $Z_2\times Z_2$ in $\tilde{H}$. Note that Oshikawa found that the nonlocal unitary can be written as~\cite{Oshikawa}
\be
U=\prod_{j<k}e^{i\pi S_j^z S_k^x}.\ee
This string order parameter is now understood as one of the order parameters to detect nontrivial SPT order in one dimension~\cite{PollmannTurner} and it can be computed using the MPS representation and examining how local tensors are transformed, along the line discussed in  Sec.~\ref{sec:SPT}.1.
\subsection{Hidden cluster order}
There is another kind of hidden order in AKLT states. Consider local projectors of rank-2: $F_\alpha=|S_\alpha=S\rangle\langle S_\alpha=S|+|S_\alpha=-S\rangle\langle S_\alpha=-S|$. It was shown that AKLT state can be converted to the so-called cluster state~\cite{RB01b} by the action of appropriate local projector~\cite{Wei2011},
\begin{equation}
    |\psi_{\rm cluster}\rangle=c\prod_{v\in A}\prod_{u\in B} F_{x}^{[v]} F_{z}^{[u]} |\psi_{\rm AKLT}\rangle,
\end{equation}
 where $c$ is a normalization constant and the effective qubit is defined by the two levels $|S_x=+S\rangle$ and $|S_x=-S\rangle$ on the $A$ sublattice and $|S_z=+S\rangle$ and $|S_z=-S\rangle$ on the $B$ sublattice. In fact, one can place $F$'s arbitrarily and randomly on a bipartite lattice that hosts the AKLT model and this will convert the AKLT state to some random graph state and this was used in the measurement-based quantum computation with AKLT states~\cite{Wei2011}; see also below in Sec.~\ref{sec:MBQC}. 

\subsection{Hidden frustration on frustrated lattices}  If the lattice is not bipartite, there is some frustration that is only revealed by considering these $F$'s. It turns out that on any loop with odd number of sites, one cannot place $F_\alpha$ with the same label $\alpha$ along such a loop. Product of such operators will annihilate the AKLT state~\cite{Wei2013}. This is due to the singlet construction of the AKLT wave function and the frustration of antiferromagnetism on such a loop. The simplest example is a periodic three-site spin-1 chain (i.e. a triangle). Another nontrivial example is the triangles on the star lattice, which hosts a spin-3/2 AKLT state. 

\section{Applications in quantum computation}
\label{sec:MBQC}

AKLT states play a role in a scheme of universal quantum computation, so-called measurement based quantum computation (MBQC) \cite{RB01}. Therein, the process of quantum computation is driven by local measurements; no unitary evolution ever takes place. The computation begins in an appropriately entangled state such as a cluster state. The cluster state is to MBQC what a blank sheet of paper is to the artist: a great number of possibilities. Every quantum circuit can be imprinted on it by the local measurements. Because of this property, cluster states are {\em{universal}} resources for measurement based quantum computation.

One may now ask: Are cluster states the only universal resource states? If not, how rare are universal resource states in Hilbert space?---Both questions have in fact been answered. Universal resource states are very rare \cite{Eisert}, but the cluster states are not the only ones. In fact, AKLT states on various lattices in 2D are universal \cite{Miyake2011,Wei2011,Wei2013,Wei2014,Wei2015,Wei2018}. The purpose of this section is to explain why this is so. To prepare for the main argument, we review the simpler but non-universal case of one dimension first.

\subsection{One dimension} 

From the perspective quantum computation, our main interest is in the two-dimensional  scenario---quantum computationally universal. However, the basic techniques that MBQC on AKLT states employs are easier to understand in 1D. We therefore start out with the one-dimensional case.

\subsubsection{Logical identity and one-qubit gates}

We explain the computational power of the AKLT chain in terms of its matrix product representation. This method is intuitive and has the added benefit of preparing for the stronger result that the MBQC power found at the AKLT point is uniform across the entire surrounding symmetry protected phase with $\mathbb{Z}_2\times \mathbb{Z}_2$-symmetry~\cite{Raussendorf2017}. 

We have derived earlier the matrix-product form for the 1D spin-1 AKLT state, e.g. in Eq.~\eqref{eqn:1Dopen}.
If we define a different orthonormal basis for the spin-1 states via\[
|0\rangle\equiv|z\rangle, \, |+1\rangle \equiv -(|x\rangle +i |y\rangle)/\sqrt{2}, \ |-1\rangle\equiv (|x\rangle -i|y\rangle)/\sqrt{2}
\]
The we obtain the matrix-product representation in this new basis:
\[P_v
\left(\begin{array}{cc}
|10\rangle & |11\rangle \\
-|00\rangle & -|01\rangle\end{array} \right)=\frac{1}{\sqrt{2}}(|x\rangle \sigma_x + |y\rangle \sigma_y + |z\rangle \sigma_z)\]
Thus measuring the physical degrees of freedom in the orthonormal basis ${\cal{B}}_0=\{|x\rangle,|y\rangle,|z\rangle\}$ has the effect of applying sequences of Pauli matrices on the virtual space. The ``virtual'' quantum register, initialized in the state described by the right boundary condition $v_R$, is thus propagated across the chain~\cite{GrossEisert},
\begin{equation}
v_R \rightarrow \sigma_{\alpha_N}v_R \rightarrow \sigma_{\alpha_{N-1}} \sigma_{\alpha_N}v_R \dots \rightarrow \prod_{i=1}^N \sigma_{\alpha_{i}} v_R.
\end{equation}
This is the simplest conceivable quantum protocol -- a quantum wire. The Pauli matrices applied to the virtual quantum register are random but known, as in quantum teleportation.

To progress from wire to logical quantum gates one simply changes the measurement basis. For example, a measurement in the basis
$$
{\cal{B}}(\phi) =\{|x_\phi\rangle = \cos\phi |x\rangle + \sin\phi |y\rangle,\, |y_\phi\rangle = -\sin \phi |x\rangle + \cos\phi |y\rangle,\, |z\rangle\} 
$$
produces a logical gate
$$
U(\phi)=\left\{
\begin{array}{rl} 
\sigma_x e^{i\phi \sigma_z},& \text{if outcome } x_\phi,\\ 
\sigma_y e^{i\phi \sigma_z},& \text{if outcome } y_\phi,\\ 
\sigma_z, & \text{if outcome } z. 
\end{array} \right.
$$
This is a probabilistic heralded rotation about the $z$-axis. When the outcome $z$ is obtained, the gate fails; but it can be reattempted as often as needed. By similar deviations from the basis ${\cal{B}}(0)$, logical rotations about the $x$ or $y$-axis can be realized, together forming a one-qubit universal set of quantum gates.

It has been observed by Else et al.~\cite{Else2012} that the capability for wire is not only a property of the AKLT state, but instead of the entire Haldane phase. It is a manifestation of symmetry protected topological order. Namely, the generic state in the Haldane phase can be written in the MPS form with
\begin{equation}\label{factor}
A_\alpha= \sigma_\alpha \otimes B_\alpha,
\end{equation}
where $B_\alpha$'s are not fixed by symmetry. One may now divide the correlation space into a logical system on which the Pauli operators act, and a ``junk subsystem'' on which the unknown matrices $B_\alpha$ act. The former supports wire as before, and the latter is simply not used.

As it turns out, the capability to enact logical gates also extends beyond the AKLT state to the entire surrounding SPT phase with $\mathbb{Z}_2\times \mathbb{Z}_2$ symmetry. See~\cite{Raussendorf2017,Stephen2017} for the techniques by which this is accomplished; here we merely point out what the additional difficulty is. Namely, to do more than wire, the spins in the chain need to be measured in bases other than ${\cal{B}}(0)$. But then the factorization property Eq.~(\ref{factor}) no longer holds. As a consequence, the logical and the junk subsystem become entangled, which leads to decoherence on the logical subsystem. This decoherence must be very carefully managed.

\subsubsection{Reduction to the 1D cluster state}\label{Reduc}

We now provide a second proof of the usefulness of 1D AKLT states as computational resources in measurement-based quantum computation, by mapping them to 1D cluster states under local operations. It is this argument that will generalize to lattice dimension two, and also to higher spins.

To simplify the discussion, we consider both the 1D AKLT state and the 1D cluster state on rings rather than chains. This does not affect quantum computational power.

The one-dimensional AKLT state can be understood within the valence bond picture, as  illustrated in Fig.~\ref{AKLT} (a). Therein, the spin-1 particle at each site $v$ on a ring is viewed as a pair of virtual spin-1/2 particles, or qubits, to which a projection $P_v$ onto the spin-1 subspace is applied. The projector takes the explicit form
$$
P = |S_z=1\rangle \langle 00| + |S_z=-1\rangle \langle 11| + |S_z=0\rangle \langle \psi^+|,
$$
where $|\psi^+\rangle = (|01\rangle +|10\rangle)/\sqrt{2}$, and $|S_z\rangle$ denote the eigenstates of the $z$-component $\hat{S}_z$  of the spin operator. The virtual spin-1/2 particles form spin singlets (Bell states) $|\psi^-\rangle =(|01\rangle - |10\rangle)/\sqrt{2}$ between neighbouring sites. Denoting the edges between neighbouring sites by $e$, the 1D AKLT state thus takes the form
\begin{equation}
|\text{AKLT}_{1D}\rangle = \bigotimes_v P_v \bigotimes_e |\psi^-\rangle_e.
\end{equation} 
Our task is to convert this AKLT state to the 1D cluster state by local operations. The latter is already known to be a computational resource \cite{RB01}. 

\begin{figure}
\begin{center}
\includegraphics[width=0.95\textwidth]{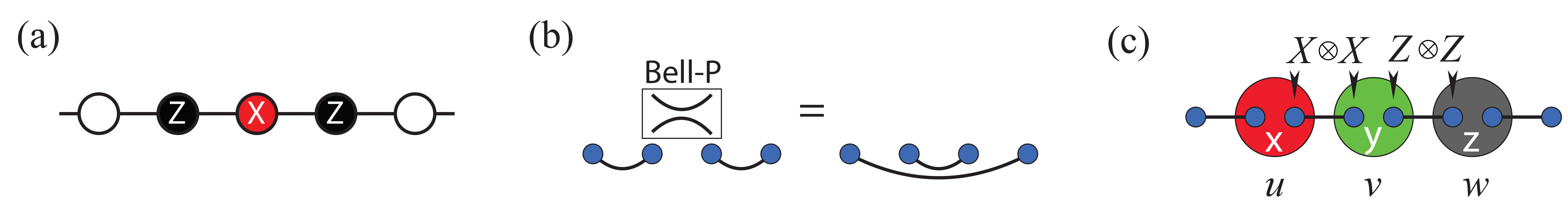}
\caption{\label{AKLT1Dfig} Conversion of the AKLT state to the cluster state in 1D. (a) 1D cluster state with stabilizer generator. (b) Bell measurement on the virtual qubits of a given site leads to entanglement swapping. (c) Element of the stabilizer of the state $|\tilde{\Psi}(\{s_v\})\rangle$. For explanation see text.}
\end{center}
\end{figure}

The 1D cluster state is a multi-qubit state, one qubit on each site $i=1,..,N$ of a ring. Up to a global phase, the cluster state is uniquely specified by the stabilizer relations
\begin{equation}\label{Clus}
|{\cal{C}}_{1D}\rangle = Z_{i-1}X_iZ_{i+1} \, |{\cal{C}}_{1D}\rangle, 
\end{equation}
where here and below $X\equiv \sigma_x$, $Y \equiv \sigma_y$ and $Z\equiv \sigma_z$.

The local reduction from the AKLT state to the cluster state proceeds by a generalized measurement, or POVM, with the three elements
\begin{equation}\label{POVMspin1}
\begin{array}{rcl}
F_z &=& \left( |S_z=1\rangle \langle S_z=1|+  |S_z=-1\rangle \langle S_z=-1|\right)/\sqrt{2},\\
F_x &=& \left( |S_x=1\rangle \langle S_x=1|+  |S_x=-1\rangle \langle S_x=-1|\right)/\sqrt{2},\\
F_y &=& \left( |S_y=1\rangle \langle S_y=1|+  |S_y=-1\rangle \langle S_y=-1|\right)/\sqrt{2}.
\end{array}
\end{equation}
These POVM elements satisfy the required completeness relation 
\begin{equation}
\sum_{\alpha \in \{x,y,z\}} F^\dagger_\alpha F_\alpha = \mathbb{I}_{S=1}.
\end{equation}
Denoting by $s_v \in \{x,y,z\}$ the POVM outcome at $v$, for all sites $v$, the post-POVM states are
\begin{equation}\label{PMS}
|\Psi(\{s_v\})\rangle:= \bigotimes_{v} F_{v,s_v}\, |\text{AKLT}_{1D}\rangle.
\end{equation}
In the following, we identify the spin-1 Hilbert space with the symmetric subspace of the pairs of the virtual spin-1/2 particles, e.g., $|S_z=1\rangle =|00\rangle$, $|S_z=-1\rangle=|11\rangle$. Up to this identification, we have $F_\alpha P= \tilde{F}_\alpha$, $\alpha = x,y,z$, with
\begin{equation}\label{POVM2}
\begin{array}{rcl}
\tilde{F}_z &=& \left(|00\rangle \langle 00| +|11\rangle \langle 11|\right)/\sqrt{2},\\
\tilde{F}_x &=& \left(|++\rangle \langle ++| +|--\rangle \langle --|\right)/\sqrt{2},\\
\tilde{F}_y &=& \left(|i,i\rangle \langle i,i| +|-i,-i\rangle \langle -i,-i|\right)/\sqrt{2}.
\end{array}
\end{equation}
Therein, $|\pm\rangle = (|0\rangle \pm |1\rangle)/\sqrt{2}$ and $|\pm i\rangle = (|0\rangle + i|1\rangle)/\sqrt{2}$. Thus, Eq.~(\ref{PMS}) simplifies to 
\begin{equation}\label{PMS2}
|\Psi(\{s_v\})\rangle= \bigotimes_{v} \tilde{F}_{v,s_v}\,\bigotimes_e |\psi^-\rangle_e.
\end{equation}
We observe that the POVM elements $\tilde{F}_\alpha$ all have rank 2. Therefore, after the POVM Eq.~(\ref{POVMspin1}) we remain with one qubit worth of Hilbert space per site. 

We now show that, irrespective of the set of POVM outcomes $\{s_v, v=1..N\}$, the post-POVM states $|\Psi(\{s_v\})\rangle$ are indeed 1D cluster states, up to local unitary equivalence and an encoding. Related to the encoding---the precise form of which depends on the POVM outcomes $\{s_v\}$---we first need to discuss ``domains'', and how to shrink them to individual qubits. As a remark, there is actual an alternative measurement scheme that converts the 1D AKLT state into a 1D cluster state~\cite{ChenDuanJiZeng}.

\paragraph{Domains.} Ring segments of nearest-neighbouring sites on which the same POVM outcome was obtained are called ``domains''.  We extract one cluster qubit per domain, undoing the encoding mentioned above. This proceeds by measuring all but one site in each domain in the basis
$$
\begin{array}{rcll}
{\cal{B}}_z &=& \left\{|z,\pm\rangle := (|00\rangle\pm |11\rangle)/\sqrt{2}\right\} & \text{if } s = z,\\
{\cal{B}}_x &=& \left\{|x,\pm\rangle := (|++\rangle\pm |--\rangle)/\sqrt{2}\right\} & \text{if } s = x,\\
{\cal{B}}_y &=& \left\{|y,\pm\rangle := (|i,i\rangle\pm |-i,-i\rangle)/\sqrt{2}\right\} & \text{if } s = y.
\end{array}
$$
It is easily checked that all $\langle \alpha,\pm|\tilde{F}_\alpha$ are Bell states (bras). Therefore, measuring in the basis ${\cal{B}}_\alpha$ following the local POVM with outcome $s=\alpha$, amounts, for all outcome combinations, to the projection onto a Bell state. We thus implement entanglement swapping,
disentangling the measured site, and otherwise leaving the entanglement structure intact. See Fig.~\ref{AKLT1Dfig} (b) for a graphical illustration. In this way, we can eliminate all redundant sites in a domain. The result is a state very similar to that of Eq.~(\ref{PMS}), but with three differences: (a) there are fewer qubits than initially, (b) now all pairs of neighbouring sites the POVM outcomes differ, and (c) the Bell states in the PEPS representation are not necessarily spin singlets $|\psi^-\rangle$ anymore, but can be either Bell state due to the measurement outcomes invoked in the entanglement swapping. Properties (a) and (b) are important for the subsequent argument, and (c) poses no obstacle. To summarize, the state after shrinking the domains is
\begin{equation}\label{PMS3}
|\tilde{\Psi}(\{s_v\})\rangle= \bigotimes_{v} \tilde{F}_{v,s_v}\,\bigotimes_e |\text{Bell}(e)\rangle_e.
\end{equation}

\begin{table}
\begin{center}
\begin{tabular}{|l|r|r|r|} \hline
POVM outcome & $z$ & $x$ & $y$ \\ \hline \hline
stabilizer generator & $Z_{v:1}Z_{v:2}$ &  $X_{v:1}X_{v:2}$ & $Y_{v:1}Y_{v:2}$  \\
$\overline{X}$ & $X_{v:1}X_{v:2}$ & $Z_{v:1}Z_{v:2}$ &  $Z_{v:1}Z_{v:2}$\\
$\overline{Z}$ & $Z_{v:1}$ & $-X_{v:1}$ & $Y_{v:1}$\\
$\overline{Y}$ & $Y_{v:1}X_{v:2}$ & $Y_{v:1}Z_{v:2}$ & $-X_{v:1}Z_{v:2}$\\ \hline
\end{tabular}
\caption{\label{Code}Encoding of graph state qubits, resulting from the POVM Eq.~(\ref{POVMspin1}) on a 1D AKLT state. The site label ``$v:1$'' means left virtual qubit in site $v$, etc.}
\end{center}
\end{table}

\paragraph{Cluster states.} We now show that the state resulting from shrinking the domains is an encoded cluster state, with the encoding depicted in Table~\ref{Code}. We observe that $(\sigma_\alpha)_{v:1} (\sigma_\alpha)_{v:2} \tilde{F}_{v,\alpha} = \tilde{F}_{v,\alpha}$, for all $\alpha =x,y,z$; and with  Eq.~(\ref{PMS3}) the operators in the first line of Table~\ref{Code} do indeed stabilize the state $|\tilde{\Psi}(\{s_v\})\rangle$.

Now the various combinations of distinct POVM outcomes on three adjacent sites need to be considered on three consecutive sites $u,v,w$. For illustration, here we consider the POVM outcomes $(s_u=x,s_v=y,s_w=z)$. The state $\bigotimes_e |\text{Bell}(e)\rangle$ is an eigenstate of the Pauli observable $(X_{u:2}X_{v:1})(Z_{v:2}Z_{w:1})$, irrespective of the precise Bell state we find on the edges $e=(u,v)$ and $e'=(v,w)$. The latter affects only the eigenvalue $\pm1$. Since $X_{v:1}Z_{v:2}$ commutes with $\tilde{F}_{v,y}$, the state $|\tilde{\Psi}(\{s_v\})\rangle$ of Eq.~(\ref{PMS3}) is also an eigenstate state of $(X_{u:2}X_{v:1})(Z_{v:2}Z_{w:1})$. Consulting Table~\ref{Code}, we find
$$
(X_{u:2}X_{v:1})(Z_{v:2}Z_{w:1}) \cong X_{u:1}(X_{v:1} Z_{v:2})Z_{w:1}) = \overline{Z}_u \overline{Y}_v \overline{Z}_w.
$$
Therein, ``$\cong$'' means equivalent up to stabilizer. See Fig.~\ref{AKLT1Dfig} (c) for a graphical illustration.

All that remains to be considered are the other five orderings of $x,y,z$. The argument and result for them is analogous. In all cases, we find stabilizers of the form $\pm \overline{Z}_u \overline{Y}_v \overline{Z}_w$ or $\pm \overline{Z}_u \overline{X}_v \overline{Z}_w$. Thus, the state $|\tilde{\Psi}(\{s_v\})\rangle$ is, up to local $z$-rotations, a 1D cluster state as defined in Eq.~(\ref{Clus}).

\subsection{Two dimensions: universal computation}

Spin-3/2 AKLT states on a two-dimensional honeycomb lattice are universal resources for MBQC. This result has been established independently by \cite{Miyake2011} and \cite{Wei2011}. 

Here, we explain the method employed in \cite{Wei2011}. The overall strategy of quantum computation is the same as in the 1D case discussed in Section~\ref{Reduc}, namely to reduce the AKLT state to a cluster 2D state by suitable local measurements. The construction is probabilistic, with a success probability approaching unity in the thermodynamic limit. At the centre of the proof is a percolation argument that involves random planar graph states. 

We thus begin by explaining  graph states \cite{Hein}, which are a generalization of the cluster states we already discussed. Both cluster and graph states belong to the class of stabilizer states \cite{GoMa}, which are eigenstates of maximal sets of commuting Pauli operators. Specifically, a graph state $|G\rangle$ corresponding to the graph $G$ with vertex set $V(G)$ and edge set $E(G)$ is the unique stabilizer state defined by the constraints
$K_v\, |G\rangle = |G\rangle$, $\forall v\in V(G)$, with 
$$
K_v:= X_v\bigotimes_{w\in V(G)|\, (v,w)\in E(G)}Z_w.
$$
A graph state becomes a cluster state when the underlying graph $G$ is that of a lattice is some spatial dimension. Cluster states in dimension 2 are universal for MBQC.\smallskip

The universality proof \cite{Wei2011} consists of three steps: (i) The reduction of the honeycomb AKLT state to a random planar graph state by local POVM-measurement; with the resulting graph state depending on the measurement outcomes\footnote{This operation is in fact the starting point of both proofs \cite{Miyake2011} and \cite{Wei2011}.} (ii) Showing that the computational power only hinges on simple  connectivity properties of the resulting graph states, and is thus a percolation problem. (iii) Demonstrating by Monte Carlo simulation that the typical graph states resulting from initial POVM satisfy these connectivity properties. Here we only give an outline of the proof, drawing on the analogy with the 1D case described in Section~\ref{Reduc}, and pointing to differences where they arise. The complete technical argument can be found in \cite{Wei2011,Wei2012}. \smallskip

{\em{Step 1: Mapping to graph states by a POVM.}} The first operation in MBQC on spin-3/2 AKLT states defined through Eq.~(\ref{Proj}) is a generalized measurement (a POVM). One such measurement is applied on each site $v$ of the honeycomb lattice ${\cal{L}}$, and it consists of 3 rank 2 elements $F_{v,\alpha}$. We denote $|\pm 3/2,\alpha\rangle$ as the state of the highest (+) or lowest (-) magnetic quantum number, in the direction $\alpha=\{x,y,z\}$. In close analogy to the one-dimensional case of Eq.~(\ref{POVM2}), the POVM elements then are
\begin{equation}\label{POVM}
    \begin{array}{rcl}
       \tilde{F}_{v,z} &=& \displaystyle{\sqrt{\frac{2}{3}}\left( |000\rangle \langle 000| + |111\rangle \langle 111|\right)}, \\
       \tilde{F}_{v,x} &=& \displaystyle{\sqrt{\frac{2}{3}}\left( |+++\rangle \langle +++| + |---\rangle \langle ---|\right)}, \\
       \tilde{F}_{v,y} &=& \displaystyle{\sqrt{\frac{2}{3}}\left( |i,i,i\rangle \langle i,i,i| + |-i,-i,-i\rangle \langle -i,-i,-i|\right)},
    \end{array}
\end{equation}
where $|0/1\rangle$, $|\pm\rangle = (|0\rangle \pm |1\rangle)/\sqrt{2}$, $|\pm i\rangle = (|0\rangle \pm i |1\rangle)/\sqrt{2}$ are eigenstates of the Pauli operators $X$, $Y$, $Z$ respectively. 

The linear operators $\tilde{F}_{v,z}$, $\tilde{F}_{v,x}$, $\tilde{F}_{v,y}$ do indeed form a POVM on the symmetric subspace projected onto by $P_v$ (cf. Eq.~(\ref{Proj}), $\sum_{\alpha=x,y,z} \tilde{F}_{v,\alpha}^\dagger \tilde{F}_{v,\alpha} = P_v$.
Every POVM element is proportional to a projector onto a two-dimensional subspace, and the resulting state
$$
|\Psi({\cal{A}})\rangle = \bigotimes_{v\in V({\cal{L}})} \tilde{F}_{v,\alpha_v}\, |\Phi_\text{AKLT}\rangle,
$$
with ${\cal{A}}=\{\alpha_v, v\in V({\cal{L}})\}$ the measurement record, is therefore a state of qubits---one for every vertex $v$. 

\begin{figure}
    \centering
    \includegraphics[width=0.7\textwidth]{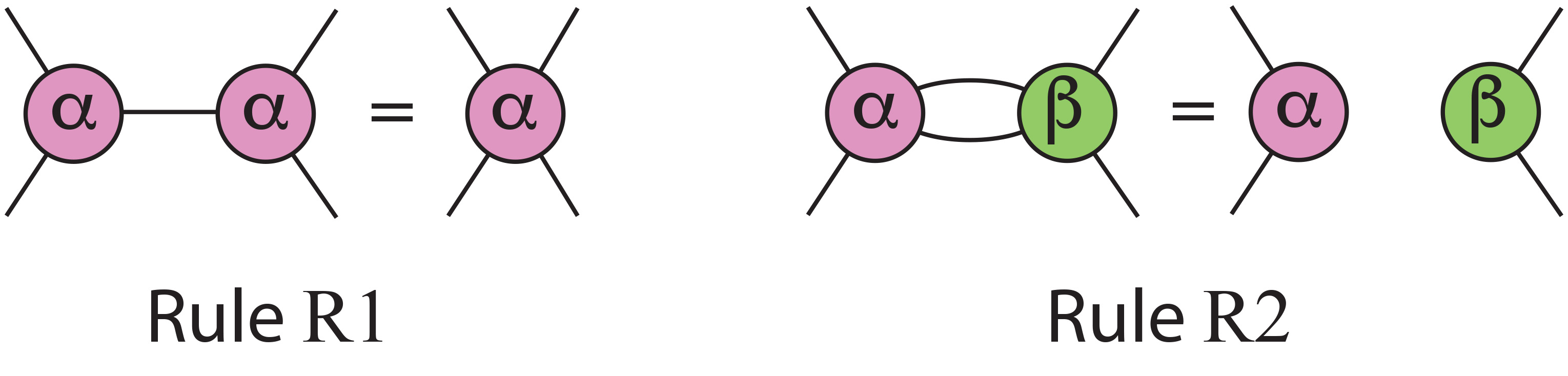}
    \caption{Graphical illustration of rules (R1) and (R2).}
    \label{R1R2}
\end{figure}

Up to local unitaries, the resulting state $|\Psi({\cal{A}})\rangle$ is an encoded graph state $|\overline{G({\cal{A}})}\rangle$, where the graph $G({\cal{A}})$ is a function of the measurement record ${\cal{A}}$. The effect of the randomness of the measurement outcomes is more severe now than it was in 1D. Like in dimension one we find domains, i.e. connected regions of lattice sites on which the same POVM outcome was obtained. As before, each domain gives rise to one encoded cluster qubit, and the encoding is undone in a similar way as before. The new feature in dimension two is that the cluster qubits resulting from the domains are connected in a random planar fashion. 
The graph $G({\cal{A}})$, depending on the random measurement record ${\cal{A}}$ and describing the resulting graph state $|G({\cal{A}})\rangle$  is obtained from the honeycomb lattice ${\cal{L}}$ and the measurement record ${\cal{A}}$ via the following two rules.
\begin{itemize}
    \item[(R1)]{[Edge contraction]: Contract all edges $e \in E({\cal{L}})$ that connect sites with the same POVM outcome.}
    \item[(R2)]{[Edge deletion]: In the resulting multigraph, delete all edges of even multiplicity, and convert edges of odd multiplicity in standard edges of multiplicity 1.}
\end{itemize}    
 For the general proof of correctness of the rules (R1) and (R2) see \cite{Wei2011}. See Fig.~\ref{R1R2} for graphical illustration.\smallskip

{\em{Step 2: the percolation problem.}} The next step is to show that the random graph state $|G({\cal{A}})\rangle$ can be converted to a standard 2D cluster state by further local measurement, if the following two conditions hold for typical graphs resulting from Step 1:
(C1) The domain size is microscopic, i.e., the size of the largest domain scales at most logarithmically with $|V({\cal{L}})|$. (C2) A left-right traversing path through $G({\cal{A}})$ exists.

Condition (C1) ensures that the graph $G({\cal{A}})$ is macroscopic if ${\cal{L}}$ is, which is required for the resulting graph state to have computational power. Condition (C2) ensures that the resulting graph states is sufficiently long-range connected. It also illustrates that we are dealing with a percolation problem.
We will comment on this observation further below.

 (C1) and (C2) are natural conditions to invoke; however, we still need to show that they are sufficient for universality. The basic argument is as follows. In the supercritical phase, where a macroscopic spanning cluster exists with high probability, this spanning cluster contains a subgraph that is topologically equivalent to a coarse-grained 2D lattice structure. Essentially, if one left-right traversing path exist, then very many such paths exist, and by symmetry of the honeycomb lattice, also very many top-bottom traversing paths exist. The corase-grained 2D subgraph can be carved out, cleaned from imperfections and finally contracted to a standard 2D grid by further local Pauli measurements on $|G({\cal{A}})\rangle$. For details see \cite{Wei2012}.
 
 We remarked above that the reduction of the AKLT state to a random planar graph state is a percolation problem, but what kind of percolation problem is it?---It resembles site percolation in so far as the random variables (POVM outcomes) live on the sites. However, it is not site percolation because no site is ever deleted. Further, the present problem resembles bond percolation in so far as edges are switched on and off (rule R2). But it is not exactly bond percolation because whether or not an edge persists is decided not simply by a probability associated with that edge. Rather it is decided by random processes associated with the nearby sites. Thus we conclude that our percolation problem defies simple characterization, and we defer its classification to further study. 
 \smallskip

{\em{Step 3: Testing the conditions (C1) and (C2).}} To complete the argument for quantum computational universality, it needs to be checked whether the typical graph state resulting from the POVM Eq.~(\ref{POVM}) satisfies the conditions (C1) and (C2). This is done numerically. 

\begin{figure}
    \centering
     (a)\\ \vspace{-0.5cm}\includegraphics[width=0.7\textwidth]{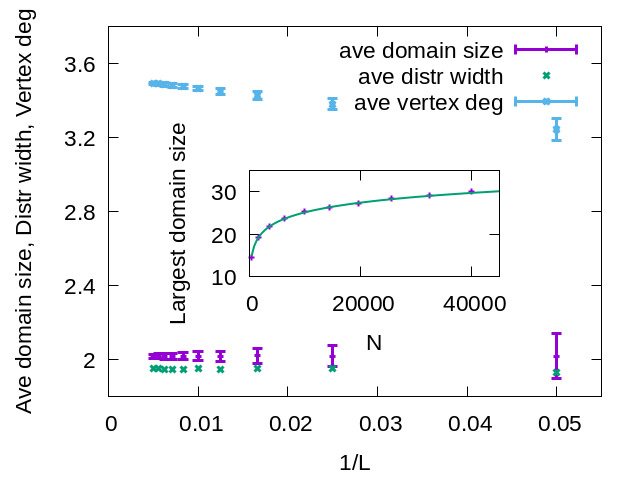}  \\
     \centering
     (b)\\ \vspace{-0.5cm}\includegraphics[width=0.7\textwidth]{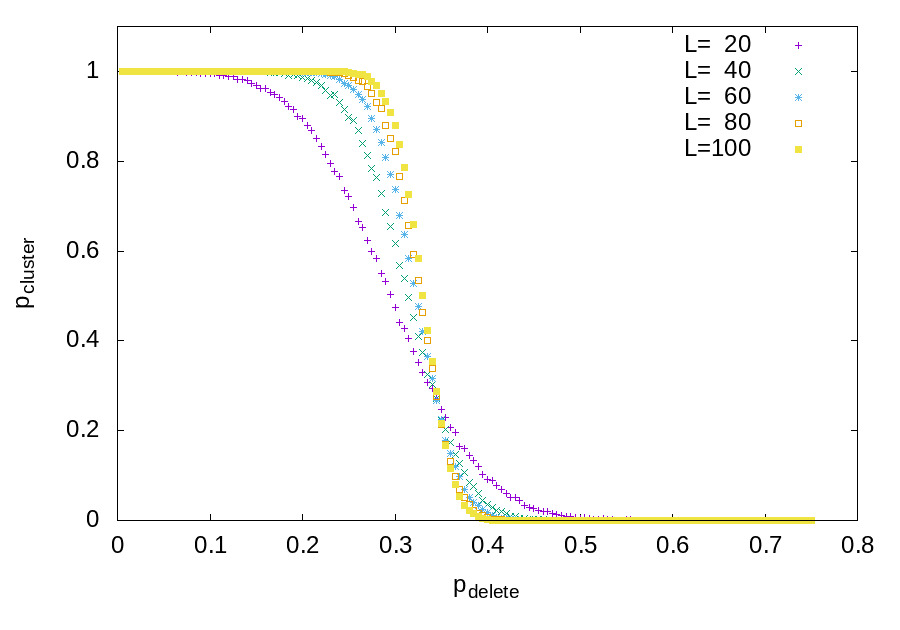}
    \caption{ (a) [Top panel] Statistics of the average domain size, average width of domain size distribution, average degree of a vertex, and the largest domain size (inset) in the typical graphs as a function of the linear size $L$. (b) [Bottom panel] Site percolation study by deleting randomly any vertex on typical random graphs and measuring the probability of a spanning cluster. The crossing represents the location of the percolation phase transition in the thermodynamic limit. Figures were reproduced from the data of the work~\cite{Wei2011}.}
    \label{MC}
\end{figure}

By rotational symmetry, for any site all three possible POVM outcomes are equally likely. However, these outcomes are correlated with outcomes on neighboring sites, and this represents a complication. For a reliable simulation, these correlations need to be taken into account. Fortunately, the joint probability for any given configuration ${\cal{A}}$ of POVM outcomes on all sites can be efficiently calculated exactly \cite{Wei2011}. Monte-Carlo simulation is thus viable, and the results are shown in Fig.~\ref{MC}. Conditions (C1) and (C2) are satisfied. This concludes the argument for computational universality of spin-3/2 AKLT states.\medskip

We conclude this section with a brief description of related work on ground states of lattice Hamiltonians as MBQC resources, and the role of symmetry. First, the argument above has been generalized to AKLT states on lattices other than honeycomb, including spin-2 \cite{Wei2013,Wei2014,Wei2015}. For most lattices, but not all, computational universality persists.

Furthermore, for a one-dimensional manifold of deformed AKLT Hamiltonians with reduced symmetry, the known transition from disorder to Ne{\'e}l order \cite{NiggemannKlumperZittartz1997} was re-investigated. It was found numerically that the location of the physical phase transition coincides with the transition in computational power~\cite{Darma}. This gave early support to the notion of ``computational phase of quantum matter'', which refers to the property of certain quantum phases---for example symmetry protected phases---to have {\em{uniform}} computational power. In such a phase, from the viewpoint of scaling, any ground state is equally good a resource for MBQC. 

The phenomenon of computational phases of matter was conceived in~\cite{HoloScreen}, where  for a hybrid of measurement based and adiabatic quantum computation it was shown that proper operation only relies on the presence of symmetry. Detailed knowledge of the Hamiltonian or its ground state is not required. The connection with symmetry protected topological order was already recognized and emphasized in this work. Subsequently, uniformity of MBQC-power in symmetry protected phases was established for one-dimensional \cite{Else2012, MM, Raussendorf2017, Stephen2017} and two-dimensional systems. In particular, 2D computationally universal phases have been identified \cite{Raussendorf2019,Stephen2019,DevWi2018, Daniel2020}.

\section{Spectral gap for AKLT models}
\label{sec:gap}

The 1D spin-1 AKLT model was proven to be gapped in the original AKLT paper~\cite{AKLT}. The significance of the result was the first proved gapped integer-value spin chain that is both isotropic in spin symmetry and gapped in the spectrum. This models differs from the spin-1 Heisenberg spin chain by the biquardratic spin-spin interaction $(\vec{S}_i\cdot\vec{S}_{i+1})^2$.
Till now, the spectral gap of the spin-1 Heisenberg model has not been rigorously established, despite accurate numerics from DMRG. The proof in the original work of AKLT analyzes in detail the ground spaces in successively increasing regions, including those being ground states in a smaller region but orthogonal to the  ground space in a larger region. In the end they were able to upper bound the projector $P_L$ to the complement of the ground space in a whole chain of size $L$ by some additive constant and a term proportional to the total Hamiltonian~\cite{AKLT}, which we quote here,
\be
P_L\le 16(l+1)\epsilon(l) + \left(\frac{2(l+1)}{e_{l+1}}+\frac{1}{e_l}\right) H_{1,L},\ee
where $e_l$ is the gap of the chain with size $l$ and $\epsilon(l)$ is an exponentially small quantity, i.e. $\epsilon(l)\le c\cdot 3^{-l}$. The finite gap exists as $16(l+1)\epsilon(l)$ can be made smaller than 1 as long as $l$ is sufficiently large. The infinite chain result was generalized from the finite chain by considering a chain with sites from $-L$ to $L$ and taking $L\rightarrow \infty$.

The technique of proving the gap in 1D had also been generalized by  Knabe~\cite{Knabe} and by Fannes, Natchtergaele, and Werner~\cite{Fannes1992}. There were also more recent works~\cite{Gosset2016,LemmMozgunov2019,Lemm2020finite}, not necessarily limited to 1D. Instead of directly bounding the Hamiltonian $H$, many of the latter developments consider bounding $H^2$, and we will discuss two variations below, which also apply to two dimensions.

Beyond one dimension, the correlation functions with respect to the ground state wave function from the hexagonal and square lattices were shown to decay exponentially~\cite{AKLT,KLT}, suggesting that the models are gapped. There were some prior numerics with tensor network ~\cite{Numerical1,Numerical2} that estimate the gap values in the thermodynamic limit. Several 2D AKLT models were recently shown to be gapped rigorously~\cite{Abdul,PomataWei2019,PomataWei2020,LemmSandvikWang2020} and 
one breakthrough came from the work of Abdul-Rahman et al. on decorated hexagonal lattices~\cite{Abdul}, where a certain number $n$ of spin 1 sites are added to each edge of the hexagonal lattice; see e.g. Fig.~\ref{fig:Decorated}b. For $n\ge 4$, they showed analytically that the decorated hexagonal lattices host AKLT models that possess a finite gap. This analytic result was generalized to other decorated lattices~\cite{PomataWei2019}, including decorated square lattices and beyond two dimensions; see e.g. Fig.~\ref{fig:Decorated}c.


AKLT Hamiltonians belong to the so-called frustration-free models, for which the ground state satisfies the lowest energy of each local term. One can simply shifts the ground state energy to be zero for convenience (which is already the case for AKLT models by the construction of projectors),
\[H_{\rm AKLT}|\Psi_{\rm AKLT}\rangle=\tilde{H}|\Psi_{\rm AKLT}\rangle =0. \]
For the purpose of proving the spectral gap, one can also replace each local term by a projector. 
We can thus consider the following Hamiltonian, $\tilde{H}=\sum_{i}  \tilde{H}_{i}$, 
where $\tilde{H}_i^2=\tilde{H}_i$ and $\tilde{H}_i|GS\rangle=0$.
If one can show that $\tilde{H}^2> \epsilon \tilde{H}$ for $\epsilon>0$, then $\tilde{H}$ has a nonzero gap (at least $\epsilon$) above the ground state(s). Thus one squares the Hamiltonian:
\begin{eqnarray}\label{eqn:Hsquare}
 (\tilde{H})^2&=&\sum_i\tilde{H}_{i}+\sum_{i\ne j}\tilde{H}_{i}\tilde{H}_{j} =\tilde{H} + \underbrace{\sum_{i\&j\, \text{overlap}} \tilde{H}_{i}\tilde{H}_{j}}_{Q\, \rm type} +\underbrace{ \sum_{i,j \rm\, no \,overlap} \tilde{H}_{i}\tilde{H}_{j}}_{R\, \rm type} \end{eqnarray}
There are at least two main different approaches that one can proceed from here.

\begin{figure}
\centering 
\includegraphics[width=0.75\textwidth]{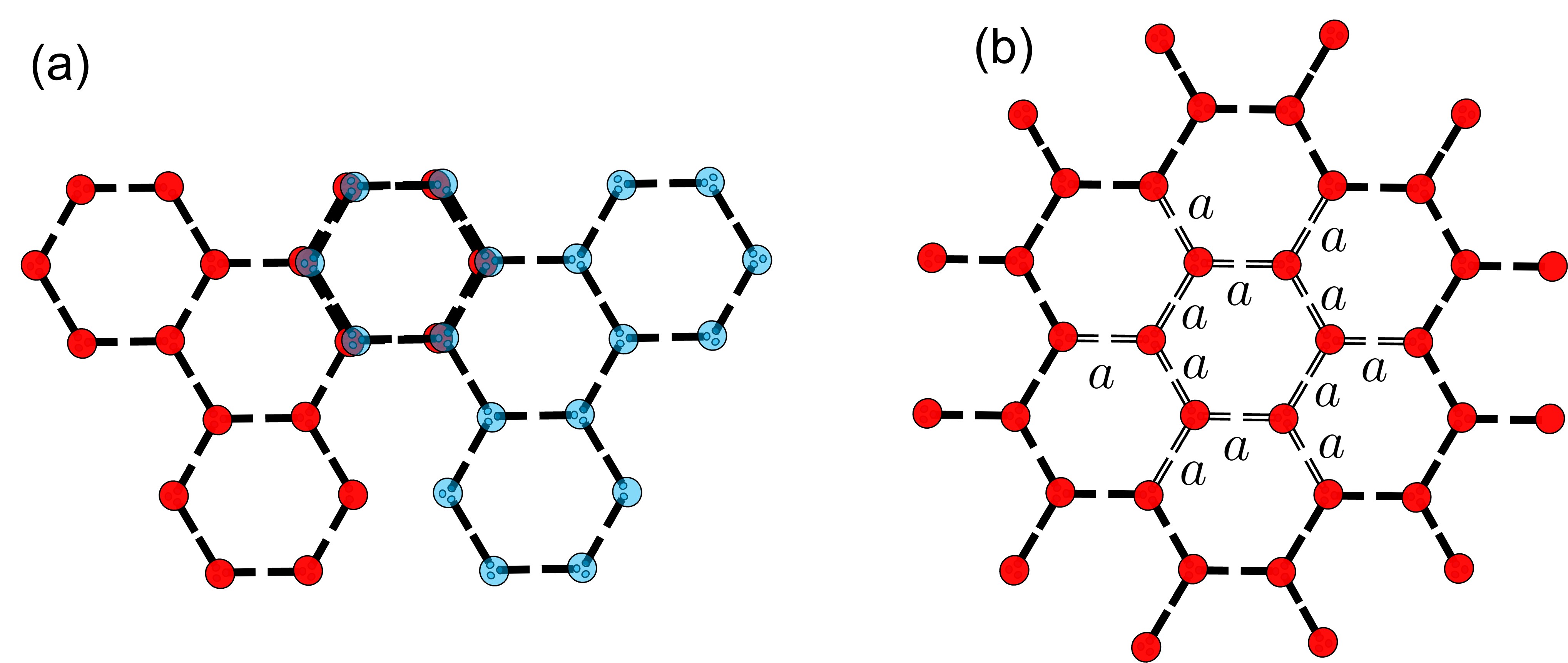}
\caption{(a) Two overlapping regions used to demonstrate the gap of the hexagonal AKLT model in Ref.~\cite{PomataWei2020}. One elementary cell consists of four hexagons, e.g., indicated by the red dots (or separately the blue dots). It overlaps with $z=6$ neighboring cells. (b) The finite-size problem of weighted AKLT model used to demonstrate the nonzero gap of the hexagonal AKLT model in Ref.~\cite{LemmSandvikWang2020}. The symbols $a$'s are used to indicate the weights of the Hamiltonian terms that are not unity. }
\label{fig:Tiling}
\end{figure}

\noindent {\bf Approach (i)}. First, the product of two non-overlapping projectors is still positive semi-definite, $\tilde{H}_i \tilde{H}_j \ge 0$ (if the two supports do not overlap), one can drop them to obtain an lower bound:
\begin{eqnarray}
 (\tilde{H})^2&
 \ge& \tilde{H} + \sum_{\langle i, j\rangle} \{\tilde{H}_{i},\tilde{H}_{j} \}. \end{eqnarray}
For the two projectors that overlap, their anticommutator $\{\tilde{H}_{i},\tilde{H}_{j} \}$ can have negative eigenvalues. However, one can also find a positive $\eta>0$ such that 
$\{\tilde{H}_{i},\tilde{H}_{j} \}\ge -\eta (\tilde{H}_{i}+\tilde{H}_{j})$ and if this $\eta$ is small enough (i.e. $\eta <1/z$, where $z$ is the coordination number), then we have
\begin{eqnarray}
 (\tilde{H})^2&
 \ge& \tilde{H}  + \sum_{\langle i, j\rangle} (\tilde{H}_{i}+\tilde{H}_{j} ) =(1-z\eta)\tilde{H}. \end{eqnarray}
 We note that it is necessary to choose $\tilde{H}_i$ to be supported nontrivially on a region larger than just nearest two neighboring sites, e.g., consecutive $n$ sites in one dimension and e.g., a few sites in small patches, which union cover all lattice sites.
This method was recently used to demonstrate the existence of the gap for AKLT models on various degree-3 lattices, in particular the hexagonal lattice, and other decorated lattices, see e.g. Fig.~\ref{fig:Decorated}, such as the singly decorated hexagonal, square and diamond lattices, as well as two other planar degree-4 lattices~\cite{PomataWei2020,GuoPomataWei2021}.   One key ingredient is to choose an appropriate tiling with an unit cell that contains a sufficiently large (but not too large) number of sites. Figure~\ref{fig:Tiling}a shows a particular choice of two unit cells and their overlap for the hexagonal lattice. There are 30 spin-3/2 sites involved, with  the Hilbert space dimension being $2^{60}$. By employing tensor-network methods, this is substantially (numerically exact) reduced to $2^{26}$, for which the computation of $\eta$ can be made with high precision. That the obtained $\eta= 0.1445124916$ is less than $1/z=1/6$ demonstrates the existence of a nonzero  gap for the AKLT model in the thermodynamic limit.

\smallskip{\noindent {\bf Approach (ii)}.}  A second method is to consider additionally a subset of terms in $\tilde{H}$ and the relation of its square to that of $\tilde{H}$. Let us first illustrate it with one dimensional model: $\tilde{H}=\sum_i\tilde{H}_{i,i+1}$ and define 
 \begin{equation}h_{n,i}=\sum_{j=i}^{n+i-1} \tilde{H}_{j,j+1},
 \end{equation}
 where $\tilde{H}_{j+N,j+1+N}=\tilde{H}_{j,j+1}$.
 Let us assume the gap of this finite system of size $n$ is $\epsilon_n$, i.e., $h_{n,i}^2\ge \epsilon_n h_{n,i}$. Equation~(\ref{eqn:Hsquare}) in this one-dimensional case becomes
 \begin{equation}
     (\tilde{H})^2=\tilde{H} +\sum_{|i-j|=1}\tilde{H}_{i,i+1}\tilde{H}_{j,j+1}+ \sum_{|i-j|>1}\tilde{H}_{i,i+1}\tilde{H}_{j,j+1}.
 \end{equation}
 We will seek to lower bound it in the form
 \begin{equation}
     (\tilde{H})^2\ge \alpha \sum_{i=1}^N (h_{n,i})^2 -\beta \tilde{H}.
 \end{equation}
 By inspection, we find that the choice with $\alpha=1/(n-1)$ and $\beta=1/(n-1)$ works. This leads to
 \begin{equation}
     (\tilde{H})^2\ge \Big(\frac{n\epsilon_n}{n-1} - \frac{1}{n-1}\Big) \tilde{H}\ge \frac{n}{n-1}\big(\epsilon_n-\frac{1}{n}\big)\tilde{H}. 
 \end{equation}
 If the finite-size gap $\epsilon_n$ is greater than $1/n$, the system with periodic boundary condition is gapped for any size greater than $n$. Knabe calculated that $\epsilon_4=0.3333> 1/4$ and thus by checking just a simple four-site problem the existence of a nonzero gap in the 1D spin-1 AKLT chain is established. This method by Knabe~\cite{Knabe} can be generalized to two dimensions and one can even allow projectors in the $n$-size unit cell ${\cal F}$ to have different weights, e.g. $h_{\cal F}=\sum_{j\in {\cal F}} w_j \tilde{H}_j$ with a gap $\gamma_F(\{w\})$.  
 
 One then considers $A=\sum_{\cal F} h_F^2$ and can derive two relations~\cite{LemmSandvikWang2020},
 \begin{eqnarray}
 A&\ge& f(\{w\})\gamma_F(\{w\}) \tilde{H}, \\
 A&\le&f(\{w^2\}) \tilde{H} + g(\{w\}) (Q+R).
 \end{eqnarray}
 From these, one obtains that
 \begin{equation}
     (\tilde{H})^2\ge \frac{f(\{w\})}{g(\{w\})}\left(\gamma_F(\{w\}) - \frac{f(\{w^2\})-g(\{w\})}{f(\{w\})}\right) \tilde{H}.
 \end{equation}
 One has the freedom to adjust the positive weights $w_j$'s and if the finite-size gap $\gamma_F(\{w\})$ for such a choice of weights is larger than the threshold
 \begin{equation}
     \Delta_{\rm TH}(\{w\})\equiv\frac{f(\{w^2\})-g(\{w\})}{f(\{w\})},
 \end{equation}
 then the Hamiltonian $\tilde{H}$ is gapped. By using this latter approach and DMRG numerical method for computing the finite-size gap, such as that shown in Fig.~\ref{fig:Tiling}b, Lemm, Sandvik and Wang showed the existence of a gap for the honeycomb-lattice AKLT model~\cite{LemmSandvikWang2020}. The Numerical DMRG method was  used to compute the finite-size gap for the problem involving 36 spin-3/2 sites and they found that the numerically obtained gap at $a=1.4$ is $\gamma_F(a=1.4)\approx  0.14599$, within sufficient accuracy, being greater than the threshold
  $\Delta_{\rm TH}(a=1.4)= 0.138$. This demonstrates that the AKLT model on the hexagonal lattice is gapped in the thermodynamic limit. 
 
Let us mention some numerical estimates of the gap value for a few AKLT models: $\Delta_{\rm 1D}\approx 0.350$, $\Delta_{\rm Hex}\approx0.10$, $\Delta_{\rm Sq}\approx 0.015$~\cite{Numerical1,Numerical2}.
 Both kinds of approaches described above have been successfully applied to showing the existence of the gap in the particular AKLT model on the honeycomb lattice. However, rigorous establishment of the gap on the square-lattice AKLT model is  still missing.

\section{
Deformed AKLT models and phase transitions}
\label{sec:deformed}
In this section, we describe examples beyond the original AKLT models by certain form of deformation. We will first discuss a one-dimensional example and then describe a few two-dimensional deformed models.
\begin{figure}
\centering 
\includegraphics[width=0.9\textwidth]{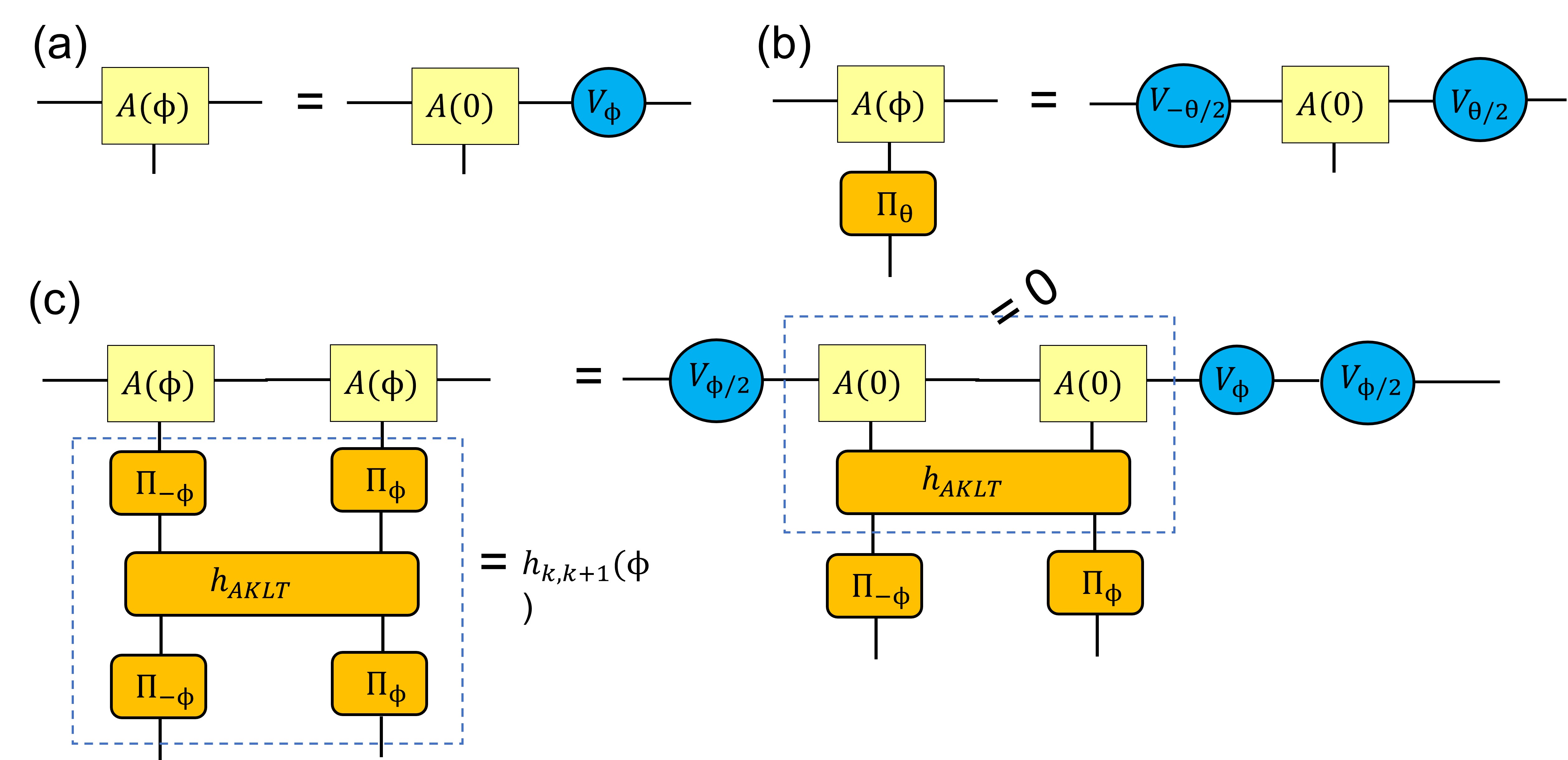}
\caption{1D deformed AKLT state: (a) relation of the local matrix to that of the undeformed AKLT state, (b) symmetry of the matrix under the action of $\Pi_\phi$, and (c) the illustration of how the individual term in the Hamiltonian annihilates the local two-site MPS. We use $h_{\rm AKLT}$ to denote a  Hamiltonian term $P_{k,k+1}^{S=2}$ in the spin-1 AKLT chain.
\label{fig:1Ddeformed}}
\end{figure}
\subsection{1D deformed AKLT chain}
In a work by Verstraete, Mart\'in-Delgado, and Cirac~\cite{VerstraeteMartin-DelgadoCirac}, they consider deforming the 1D AKLT Hamiltonian,
\begin{equation}
    H_{\rm VMC}(\phi)=\sum_k h_{k,k+1}(\phi)=\sum_k (\Pi_\phi)^{-1}\otimes \Pi_\phi\, P^{S=2}_{k,k+1}\,  (\Pi_\phi)^{-1}\otimes \Pi_\phi,
\end{equation}
where 
\begin{equation}
  \Pi_\phi=\left(\begin{array}{ccc}
e^{\phi} & 0 & 0 \\
0 & 1 & 0\\
0 & 0 & e^{-\phi} \end{array} \right).
\end{equation}
The ground state can also be represented by a translation-invariant MPS, via
$A_s(\phi)=A_s^{\rm AKLT}V_\phi$,
where $A_s^{\rm AKLT}$'s are matrices shown in Eq.~\eqref{eqn:1DMPS} and
\begin{equation}
   V_\phi\equiv \left(\begin{array}{cc}e^{\phi} & 0 \\0 & e^{-\phi}\end{array} \right).
\end{equation}
The point $\phi=0$ is the original 1D AKLT chain.
The action $\Pi_\theta$ on a local spin can be translated  to that on the virtual degrees of freedom, \begin{equation}
    \sum_{s'}(\Pi_\theta)_{ss'}A_{s'}(\phi)= V_{-\theta/2} A_s(\phi) V_{\theta/2},
\end{equation}
which represents a symmetry in the MPS.
We can easily see that $h_{k,k+1}(\phi)$ annihilates $\sum_{s,s'} A_{s}A_{s'}|s,s'\rangle_{k,k+1}$ and thus the claim of the MPS represents the ground state of $H_{\rm VMC}(\phi)$ is verified; see Fig.~\ref{fig:1Ddeformed}. 
As $\phi\rightarrow\pm\infty$, the ground state of $H_{\rm VMC}(\phi)$ is a product state $|..000..\rangle$, having zero correlation length and zero entanglement length. As $\phi$ decreases its magnitude towards 0 (which represents the AKLT state), the correlation increases and reaches the maximum at the AKLT point ($\phi=0$). However, the so-called entanglement length~\cite{VerstraetePoppCirac}, i.e. the largest distance between any two sites that entanglement can be concentrated via measurement on all other sites, increases and approaches infinity at $\phi=0$. The deformed model does not possess any conventional phase transition but has a transition in the localizable entanglement~\cite{VerstraeteMartin-DelgadoCirac}.

\subsection{2D deformed AKLT models and their phase transitions}
\begin{figure}
\centering 
\includegraphics[width=0.8\textwidth]{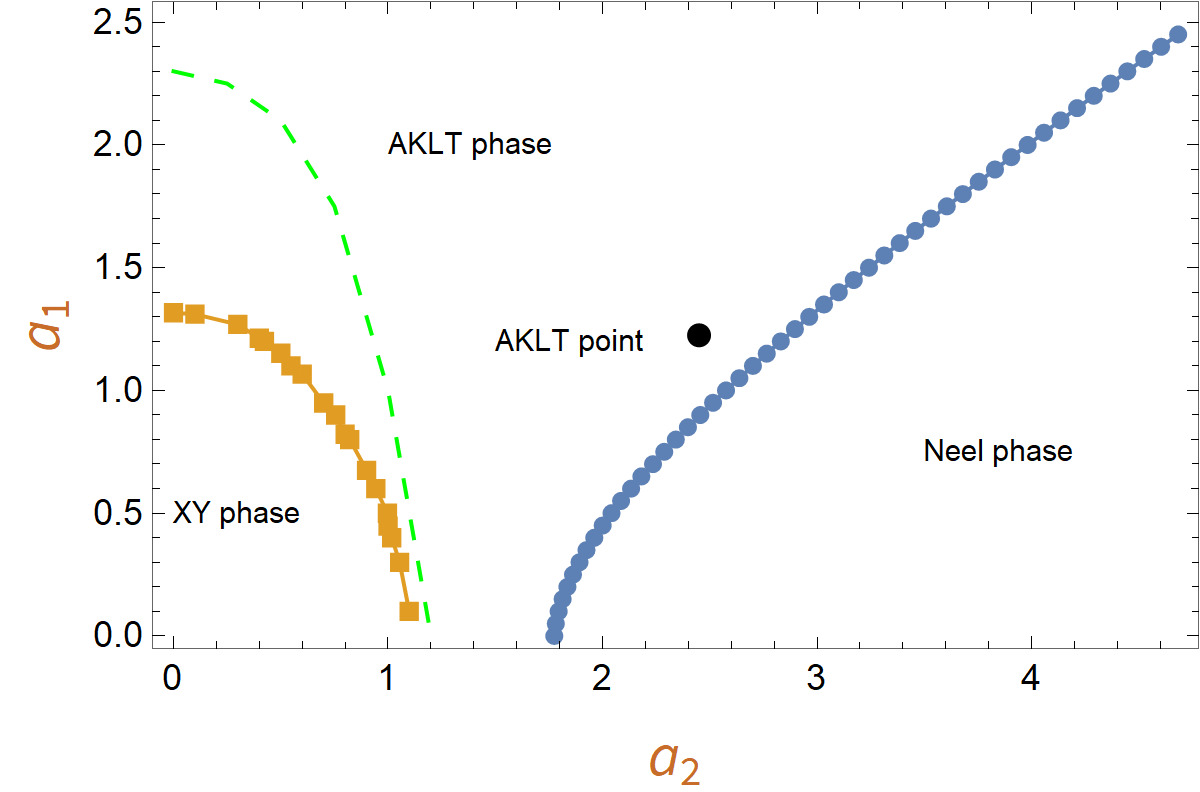}
\caption{Phase diagram of deformed AKLT model on the square lattice, reproduced from the data of the work~\cite{PomataHuangWei2018}. In the region between green dashed line and the XY phase, there is a large correlation length.  But there is no phase transition across the green dashed line.}
\label{fig:squarephase}
\end{figure}

Niggemann, A. Klümper, and J. Zittartz consider deformation from the original AKLT states on hexagonal and square lattices~\cite{NiggemannKlumperZittartz1997,NiggemannKlumperZittartz2000} and they find that using an approximate mapping to classical vertex models, there is a transition to an Ne\'el ordered state. 
Numerics using tensor-network methods also confirm this~\cite{HieidaOkunishiAkutsu99,HuangWagnerWei16,PomataHuangWei2018}. In the case of the square lattice, the deformation such that the weights of $|S=2,S_z=\pm 2\rangle$ and $|S=2,S_z=\pm 1\rangle$ are small relative to that of $|S=2,S_z=0\rangle$ gives rise to an XY phase~\cite{PomataHuangWei2018}, which was unexpected.

\smallskip\noindent {\bf Hexagonal and other trivalent lattices}. The deformation Niggemann, A. Klümper, and J. Zittartz found of the AKLT state on trivalent lattices, including the hexagonal lattice, can be achieved by the following operator, 
\begin{equation}
    \Pi^{S=3/2}(a)=\frac{a}{\sqrt{3}}\big(|+3/2\rangle\langle +3/2| + |-3/2\rangle\langle -3/2|\big) + \big(|+1/2\rangle\langle+1/2|+|-1/2\rangle\langle-1/2|\big),
\end{equation}
and applying this operator on all sites to the AKLT state gives rise to the following deformed wave function,
\begin{equation}
    |\psi_{\rm deformed}(a)\rangle\sim \Pi^{S=3/2}(a)^{\otimes N}|\psi_{\rm AKLT}\rangle.
\end{equation}
The coefficients $(a/\sqrt{3},a/\sqrt{3},1,1)$ correspond to local re-scaling of the wavefunction on $S_z=\pm3/2$ and $S_z=\pm1/2$, respectively.
They also constructed a (5-parameter) family of parent Hamiltonians such that $|\psi_{\rm deformed}(a)\rangle$
is the ground state. We refer the readers to their paper for the details of the Hamiltonians~\cite{NiggemannKlumperZittartz1997}. As far as the ground states are concerned, we can define the parent Hamiltonian as (when $a\ne 0$)
\begin{equation}
    H_{\rm deformed}^{[S=3/2]}(a)= \sum_{\langle i,j\rangle} \Pi^{S=3/2}(a)_i^{-1}\otimes \Pi^{S=3/2}(a)_j^{-1}
    \,h_{ij}^{\rm AKLT}\,
    \Pi^{S=3/2}(a)_i^{-1}\otimes \Pi^{S=3/2}(a)_j^{-1},
\end{equation}
in a way similar to the deformed Hamiltonian in one dimension.

By approximating the norm square $\langle \psi_{\rm deformed}(a)|\psi_{\rm deformed}(a)\rangle$ to a classical 8-vertex model, Niggemann, A. Klümper, and J. Zittartz were able to show that the weights of the vertex model satisfy the free-fermion condition and thus  there is an Ising-type transition at $a_c=\sqrt{3+ \sqrt{
12}}\approx 2.54$ from the valence-bond solid phase to a N\'eel phase  $a$ increases. The existence of the N\'eel order at large $a$'s can be understood easily, as in this limit local $S_z$ components are dominated by $S_z=\pm3/2$ and due to the singlet construction in the AKLT wavefunction, the neighboring sites cannot share the same $S_z$ value, hence there is the N\'eel order.  The transition was later confirmed by Huang, Wagner and Wei~\cite{HuangWagnerWei16} using a tensor-network method without the approximation used by Niggemann, A. Klümper, and J. Zittartzto a vertex model. 

Similar consideration was applied to the square-octagon lattice (still a spin-3/2 model) and Niggemann and Zittartz~\cite{NiggemannZittartz} used an 8-vertex model analysis and found the VBS-N\'eel transition at $a_c  \approx 2.65158$. The tensor network methods by Huang, Wagner and Wei yield some improvement at $a_c\approx  2.6547 $, and they also found a   different  vertex model that gives a value close to $2.6547$. They additionally discussed other trivalent lattices, such as the cross or star lattices and  calculated spontaneous magnetization.

We note that construction of AKLT states via spin triplet valence bonds can also be used and their deformation can be considered. On bipartite lattices, models with other valence bonds are equivalent under local transformations, but those that are not bipartite can have different phase diagrams under deformation. For example, using the two types of triplet $|\phi^\pm\rangle=(|00\rangle\pm |11\rangle)/\sqrt{2}$ on the star lattice, as the deformation parameter $a$ varies, there is a ferromagnetic phase for $a\le a_{c1}\approx0.5850$, a VBS phase for $a_{c1}\le a \le a_{c2} \approx 3.0243$ and another ferromagnetic phase for $a\ge a_{c2}$ for the deformed AKLT model on the star lattice~\cite{HuangWagnerWei16}. The two ferromagnetic phases differ in the axis of magnetization, e.g.  $x$ vs. $y$ axis for different triplet bonds $|\phi^\pm\rangle$.

\smallskip\noindent {\bf Square lattice}.  Different from the trivalent lattices, the AKLT model on the square lattice is spin-2. Niggemann,  Klümper, and  Zittartz consider the following deformation on the original AKLT wave function~\cite{NiggemannKlumperZittartz2000}
\begin{equation}
    D(a_1,a_2)=\frac{a_2}{\sqrt{6}}\big(|+2\rangle\langle +2| + |-2\rangle\langle -2|\big) +\frac{a_1}{\sqrt{3/2}} \big(|+\rangle\langle+1|+|-\rangle\langle-|+|0\rangle\langle0|\big).
\end{equation}
They also constructed a family of parent Hamiltonians such that 
\begin{equation}
    |\psi^{[S=2]}(a_1,a_2)\rangle\sim D(a_1,a_2)^{\otimes N}|\psi_{\rm AKLT}\rangle
\end{equation}
is the ground state. The parent Hamiltonian, as far as the ground states are concerned, can be defined as (for $a_1\ne0$ and $a_2\ne0$)
\begin{equation}
    H(a_1,a_2)^{[S=2]}= \sum_{\langle i,j\rangle} D(a_1,a_2)_i^{-1}\otimes D(a_1,a_2)_j^{-1}
    \,h_{ij}^{S=2\,\rm AKLT}\,
    D(a_1,a_2)_i^{-1}\otimes D(a_1,a_2)_j^{-1}.
\end{equation}
Using a classical vertex model and solving it via a Monte Carlo method, Niggemann,  Klümper, and  Zittartz found transitions from VBS to N\'eel phase across a transition line defined approximated by 
$a_2^2\approx 3.0 a_1^2 +3.7$.
With tensor network methods, the precise boundary between the VBS phase (which was referred to as the AKLT phase in Ref.~\cite{PomataHuangWei2018}) and the N\'eel phase was obtained. Furthermore, an XY phase was found, which is gapless and has infinite correlation. Close to the XY boundary but inside the AKLT-VBS phase, there is a region of finite but large correlation length; see Fig.~\ref{fig:squarephase}. Such a pseudo quasi-long-range region also occurs in the deformed AKLT model on the honeycomb lattice,  but there does not exist an XY phase~\cite{PomataHuangWei2018}. 

\smallskip \noindent {\bf Quantum computation with deformed AKLT states}. For these states, it is also interesting to ask whether they are also useful for quantum computation away from the exact AKLT point. This was first studied by Darmawan, Brennen and Bartlett on the honeycomb case~\cite{Darma}. We have seen in Sec.~\ref{sec:MBQC} that POVMs $\{F_x^\dagger F_x, F_y^\dagger F_y, F_z^\dagger F_z\}$ applied to all sites convert the AKLT state to a random graph that depends on the measurement outcomes. What Darmawan, Brennen and Bartlett found is essentially a modified set of POVM that undoes the  operation $\Pi^{S=3/2}(a)$ and at the same time applies the above POVM, which is possible for $a\ge 1$. However, for $a$ large enough, there is a transition to the N\'eel phase and they found that the ability for universal quantum computation disappears at this transition. This makes sense, as we do not expect a quantum N\'eel state has the entanglement necessary for MBQC. This result was generalized to other trivalent lattices with spin-3/2 AKLT states and the square lattice with a spin-2 AKLT state~\cite{HuangWagnerWei16,Wei2015}. 
\section{Conclusion}
\label{sec:conclusion}

The AKLT model was invented~\cite{AKLT} as a concrete example of Haldane's conjecture on the spectral gap of isotropic spin chains~\cite{Haldane83,Haldane83b}. The construction of the wave functions in both one- and two-dimensions was a precursor of modern matrix-product   states~\cite{Fannes1992,MPS} and projected entangled pair states~\cite{PEPS}. Their short-ranged entanglement in the presence of symmetry is also a manifestation  of symmetry-protected topological order~\cite{Gu,Pollmann,Chen}. In low dimensions such as one and two, AKLT states are disordered, possessing no local magnetization~\cite{AKLT,Param}. However, in the cubic lattice~\cite{Param} and in the Bethe lattice with a large enough coordination number~\cite{AKLT}, AKLT states display N\'eel order. In deformed AKLT models, the valence-bond solid phase can turn into a N\'eel phase as model parameters vary that locally favor maximal magnitude of $S_z$ components~\cite{NiggemannKlumperZittartz1997,NiggemannZittartz,NiggemannKlumperZittartz2000,HuangWagnerWei16}. Surprisingly, a gapless XY-like phase can emerge in such a deformed model~\cite{NiggemannKlumperZittartz2000} on the square lattice at a region where the $S_z=0$ component is locally favored~\cite{PomataHuangWei2018}. 

The 2D hexagonal AKLT model was conjectured to be gapped in the original work more than three decades ago~\cite{AKLT}. AKLT models on other lattices have also been proved, such as other degree-3 2D lattices and their decorated version lattices~\cite{Abdul,PomataWei2019,PomataWei2020,GuoPomataWei2021}. But the existence of the gap for the models on the square and kagom\'e lattices remain unproved. Interestingly, the proof was recently established with two different methods~\cite{PomataWei2020,LemmSandvikWang2020}, both utilizing techniques of tensor network, combining analytic reduction and high-precision numerics. Perhaps the most surprising aspect of AKLT states is that many of them can be used as a resource to realize universal quantum computation~\cite{Wei2011,Miyake2011,Wei2013,Wei2014,Wei2015}.  

In terms of experiments,  we mentioned earlier that the $S=1/2$ edge degrees of a Heisenberg ferromagnet was observed~\cite{edge} and confirmation of the Haldane gap were made previously~\cite{Buyers,Renard}.  A short AKLT chain was created in the photonic~\cite{Resch}   and trapped-ion systems~\cite{Senko}. Very recently, fractional excitations were observed in nanographene spin chains, which were modeled as a $S=1$ bilinear-biquadratic spin chain~\cite{FractionalEdge}, to which the AKLT chain is a special case. There are other theoretical proposals for the 1D AKLT spin, such as using measurement-induced steering on quantum spin systems~\cite{Roy} and driven-dissipative control of cold atoms in tilted optical lattices~\cite{Sharma}. Realization of two dimensional AKLT states is more challenging.
We mention that there is a theoretical work by Sela et al. on AKLT on solid-state material~\cite{Sela}. Using 2D AKLT states  for universal quantum computation may still be years ahead. But knowing that they are in principle a useful resource is intriguing, as it is a somewhat unexpected development from AKLT models.

\medskip \noindent {\bf Acknowledgments}.
T.-C.W. acknowledges  support from National Science Foundation under Grants No. PHY 1314748, No. PHY 1620252, and No.
PHY 1915165 on subjects related to AKLT models. RR is supported by the Canada First Research Excellence Fund, Quantum Materials and Future Technologies Program.

\end{document}